\documentclass[twoside,twocolumn,9pt]{article}
\usepackage{extsizes}
\usepackage[super,sort&compress,comma]{natbib} 
\usepackage[version=3]{mhchem}
\usepackage[left=1.5cm, right=1.5cm, top=1.785cm, bottom=2.0cm]{geometry}
\usepackage{balance}
\usepackage{mathptmx}
\usepackage{sectsty}
\usepackage{graphicx} 
\usepackage{lastpage}
\usepackage[format=plain,justification=justified,singlelinecheck=false,font={stretch=1.125,small,sf},labelfont=bf,labelsep=space]{caption}
\usepackage{float}
\usepackage{fancyhdr}
\usepackage{fnpos}
\usepackage{amsfonts, amsmath, amssymb}
\usepackage[english]{babel}
\addto{\captionsenglish}{%
  
}
\usepackage{array}
\usepackage{droidsans}
\usepackage{charter}
\usepackage[T1]{fontenc}
\usepackage[usenames,dvipsnames]{xcolor}
\usepackage{setspace}
\usepackage[compact]{titlesec}
\usepackage{hyperref}

\usepackage{epstopdf}

\definecolor{cream}{RGB}{222,217,201}

\begin{document}

\pagestyle{fancy}
\thispagestyle{plain}
\fancypagestyle{plain}{
\renewcommand{\headrulewidth}{0pt}
}

\makeFNbottom
\makeatletter
\renewcommand\LARGE{\@setfontsize\LARGE{15pt}{17}}
\renewcommand\Large{\@setfontsize\Large{12pt}{14}}
\renewcommand\large{\@setfontsize\large{10pt}{12}}
\renewcommand\footnotesize{\@setfontsize\footnotesize{7pt}{10}}
\makeatother

\renewcommand{\thefootnote}{\fnsymbol{footnote}}
\renewcommand\footnoterule{\vspace*{1pt}%
\color{cream}\hrule width 3.5in height 0.4pt \color{black}\vspace*{5pt}} 
\setcounter{secnumdepth}{5}

\makeatletter 
\renewcommand\@biblabel[1]{#1}            
\renewcommand\@makefntext[1]%
{\noindent\makebox[0pt][r]{\@thefnmark\,}#1}
\makeatother 
\renewcommand{\figurename}{\small{Fig.}~}
\sectionfont{\sffamily\Large}
\subsectionfont{\normalsize}
\subsubsectionfont{\bf}
\setstretch{1.125} 
\setlength{\skip\footins}{0.8cm}
\setlength{\footnotesep}{0.25cm}
\setlength{\jot}{10pt}
\titlespacing*{\section}{0pt}{4pt}{4pt}
\titlespacing*{\subsection}{0pt}{15pt}{1pt}

\fancyfoot{}
\fancyfoot[LO,RE]{\vspace{-7.1pt}\includegraphics[height=9pt]{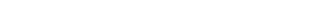}}
\fancyfoot[CO]{\vspace{-7.1pt}\hspace{13.2cm}\includegraphics{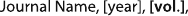}}
\fancyfoot[CE]{\vspace{-7.2pt}\hspace{-14.2cm}\includegraphics{RF}}
\fancyfoot[RO]{\footnotesize{\sffamily{1--\pageref{LastPage} ~\textbar  \hspace{2pt}\thepage}}}
\fancyfoot[LE]{\footnotesize{\sffamily{\thepage~\textbar\hspace{3.45cm} 1--\pageref{LastPage}}}}
\fancyhead{}
\renewcommand{\headrulewidth}{0pt} 
\renewcommand{\footrulewidth}{0pt}
\setlength{\arrayrulewidth}{1pt}
\setlength{\columnsep}{6.5mm}
\setlength\bibsep{1pt}

\makeatletter 
\newlength{\figrulesep} 
\setlength{\figrulesep}{0.5\textfloatsep} 

\newcommand{\topfigrule}{\vspace*{-1pt}%
\noindent{\color{cream}\rule[-\figrulesep]{\columnwidth}{1.5pt}} }

\newcommand{\botfigrule}{\vspace*{-2pt}%
\noindent{\color{cream}\rule[\figrulesep]{\columnwidth}{1.5pt}} }

\newcommand{\dblfigrule}{\vspace*{-1pt}%
\noindent{\color{cream}\rule[-\figrulesep]{\textwidth}{1.5pt}} }

\makeatother

\twocolumn[
  \begin{@twocolumnfalse}
{\includegraphics[height=30pt]{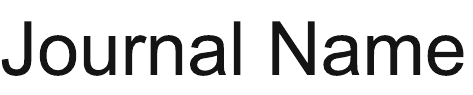}\hfill\raisebox{0pt}[0pt][0pt]{\includegraphics[height=55pt]{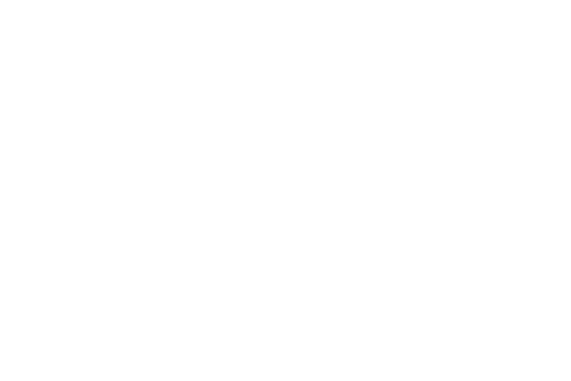}}\\[1ex]
\includegraphics[width=18.5cm]{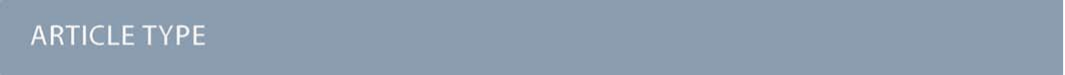}}\par
\vspace{1em}
\sffamily
\begin{tabular}{m{4.5cm} p{13.5cm} }

\includegraphics{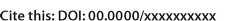} & \noindent\LARGE{\textbf{Conformation and dynamics of partially active linear polymers$^\dag$}} \\
\vspace{0.3cm} & \vspace{0.3cm} \\

& \noindent\large{Marin Vatin,$^{\ast}$\textit{$^{a,b}$} Sumanta Kundu,\textit{$^{a,b,c}$} and Emanuele Locatelli\textit{$^{a,b}$}} \\

\includegraphics{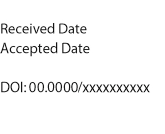} & \noindent\normalsize{We perform numerical simulations of isolated, partially active polymers, driven out-of-equilibrium by a fraction of their monomers. We show that, if the active beads are all gathered in a contiguous block, the position of the section along the chain determines the conformational and dynamical properties of the system. Notably, one can modulate the diffusion coefficient of the polymer from {active-like to passive-like} just by changing the position of the active block. Further, in special cases, enhancement of diffusion can be achieved by decreasing the overall polymer activity. Our findings may help in the modelization of active biophysical systems, such as filamentous bacteria or worms.} \\

\end{tabular}

\end{@twocolumnfalse} 
\vspace{0.6cm}
]

\renewcommand*\rmdefault{bch}\normalfont\upshape
\rmfamily
\section*{}
\vspace{-1cm}


\footnotetext{\textit{$^{a}$~Department of Physics and Astronomy, University of Padova, Via Marzolo 8, I-35131 Padova, Italy.}}
\footnotetext{\textit{$^{b}$~INFN, Sezione di Padova, Via Marzolo 8, I-35131 Padova, Italy.}}
\footnotetext{\textit{$^{c}$~International School for Advanced Studies (SISSA), 34136, Trieste, Italy.}}

\footnotetext{\dag~Electronic Supplementary Information (ESI) available: Theoretical calculation of the number of possible arrangements. Additional data on size and shape parameters, for different values of $N$ and different values of $\mathrm{Pe}$. Additional information on the end-to-end autocorrelation function. Size and shape parameters for multiple active blocks. See DOI: 00.0000/00000000.}



\section{Introduction}

The field of active matter deals with systems that move autonomously by consuming some source of fuel; this brings them out of equilibrium with respect to the surrounding environment~\cite{marchetti2013hydrodynamics}. Such a feature has sparked a lot of theoretical interest, for a few different reasons. First, the autonomous motion, also called self-propulsion, acts at the level of the individual constituents and is thus strongly different from other non-equilibrium processes. Second, the phenomenology that emerged is very rich and vastly different from equilibrium\cite{ramaswamy2010mechanics}. Third, several systems of great biological interest are active, such as molecular motors, bacteria and other micro-organisms, cells and even individuals at the macro-scale\cite{vicsek2012collective,ramaswamy2017active}. Last but not least, simple models\cite{tailleur2008statistical, ramaswamy2010mechanics, ramaswamy2017active} allowed to gain a great deal of insight on the Physics of active systems, paving the way to new experiments\cite{vizsnyiczai2017light,massana2022rectification}.\\Within active matter, active filaments are a special class, that encompasses a noteworthy spectrum of different systems, such as interphase chromatin~\cite{saintillan2018extensile}, cytoskeleton~\cite{fletcher2010cell} and actomyosin networks~\cite{koenderink2009active}, microtubule assays\cite{schaller2010polar} and, as an extension, active nematics\cite{doostmohammadi2018active}, cilia~\cite{loiseau2020active, elgeti2013emergence, chakrabarti2022multiscale}, unicellular micro-organisms~\cite{balagam2015mechanism, faluweki2022structural, patra2022collective, kurzthaler2021geometric} as well as complex, macroscopic worms~\cite{deblais2020rheology, deblais2020phase, nguyen2021emergent, patil2023ultrafast}, with recent research exploring the development of artificial systems~\cite{ozkan2021collective, zheng2023self}. Particle-based micro- and mesoscopic models for active filaments are called ``active polymers'' and come in different flavors. Indeed, one can drive the system out-of-equilibrium by imposing a different temperature to a fraction of the constituents\cite{smrek2020active}, or using a colored noise\cite{osmanovic2017dynamics} or using a self-propulsion force. Within the latter category, one can distinguish between Active Brownian Polymers\cite{kaiser2014unusual,winkler2020physics}, i.e. a collection of Active Brownian Particles with no correlation along the backbone, and Polar (or tangential) Active Polymers\cite{isele2015self,winkler2020physics}, where the direction of the self-propulsion is connected to the backbone tangent and has no internal dynamics. Notably, the latter model shows interesting properties both in three and two dimensions: it has been applied to microtubule assays\cite{schaller2010polar, sciortino2023polarity} and active nematics\cite{joshi2019interplay,vliegenthart2020filamentous,zhang2021spatiotemporal} and, in 2D, displays a rich phase diagram\cite{isele2015self, duman2018collective} as well as interesting properties at high density, e.g. the emergence of collective states characterized by topological defects\cite{prathyusha2018dynamically}.\\An intriguing application is the modelization of micro- and macro-organisms. Filamentous bacteria and other micro-organisms have indeed been modeled as active polymers: for example, it was recently shown that the collective properties of Malaria sporozoites can be rationalized by the combination of activity and polymeric properties, such as the persistence length\cite{patra2022collective}. Further, recently a polymeric model was able to rationalize the formation of a collective state observed in different worm species, called ``blob'', and formed by many entangled individuals\cite{patil2023ultrafast}. In this case, the key ingredient was provided by the peculiar pattern of the worms head; such patterns seem to be also key factors to model the motion of similar organisms\cite{krishnamurthy2023emergent}. A further application concerns the modeling of chromatin dynamics, with a focus on collective motion\cite{mahajan2022euchromatin, eshghi2023activity}\\
We consider here \textit{partially} active polymer chains which, at variance with conventional (or fully) active polymers, possess only a fraction of active monomers. In this setting, one has to choose how to arrange the active fraction $p$ along the backbone. For example, a regular pattern of active and passive beads was considered in a recent work\cite{Anand2018}, where it was shown to induce peculiar structures. Going beyond regular patterns, a completely random arrangement is the most general option. However one of the aims of this paper is to highlight the existence of a special subset of arrangements, contiguous active blocks, and to characterize their influence on the conformation and dynamics of the filaments. We will show that the distance of the active block along the polymer contour, measured from one of the ends of the polymer, strongly influences its conformation and dynamics: in a population of identical, isolated, partially active polymers, for which the contour position of the active block is placed at random along the polymer backbone, a non-zero dynamical heterogeneity emerges. This heterogeneity is not present when the active monomers are arranged completely at random.\\The inspiration for such a choice comes, similarly to other works\cite{smrek2020active,chubak2020emergence}, from chromatin: chromatin is organized, within the same gene, in ``compartments'', i.e. more or less contiguous sections with different physical and functional properties\cite{hubner2013chromatin}. In particular, ``active'' chromatin marks the active genes, whose information determines the phenotype of the cell; it is believed to be out-of-equilibrium, due to the action of the transcription machinery.\cite{chuang2006long, zidovska2013micron}\\
This paper is structured as follows: after a brief description of the polymer model employed (Sec.~\ref{sec:model}), of the simulation details (Sec.~\ref{sec:sim}) and of the definitions used throughout the paper (Sec.~\ref{sec:obs}), we first show the emergence of the dynamical heterogeneity, connected to a block arrangement of active monomers (Sec.~\ref{sec:nongauss}). Then, focusing on the specific case of a single active block, we study in detail how the contour position modifies the conformation of the chain (Sec.~\ref{sec:rgx},~\ref{sec:passive}) at its scaling exponent (Sec.~\ref{sec:scalingRg}). Moving to the dynamics, we further elucidate how the diffusion coefficient is affected by the position of the active block showing that, in a special case, the fully active behaviour can be recovered (Sec.~\ref{sec:msd},~\ref{sec:Dx}) or even enhanced by a small amount of active sites~\ref{sec:persist}. 

\section{Models and methods}

\subsection{Active polymer model}
\label{sec:model}
\begin{figure}[!h]
\centering
\begin{tabular}{c}
\includegraphics[width=1.\columnwidth]{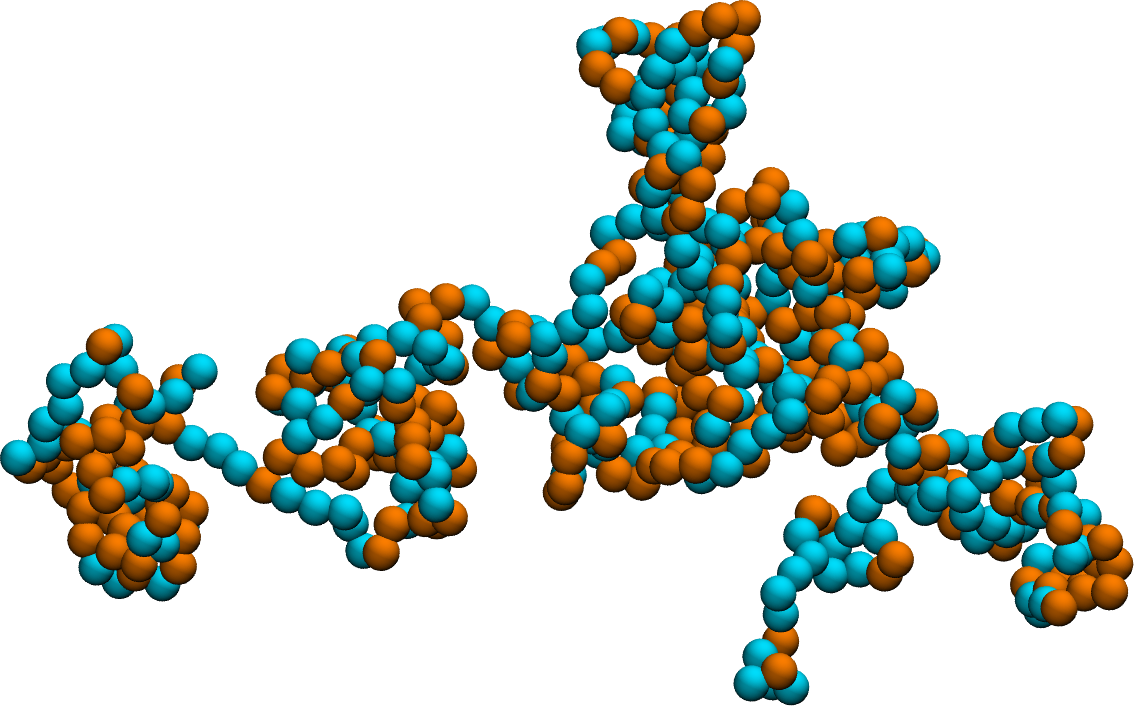} \\
(a) \\
\includegraphics[width=1.\columnwidth]{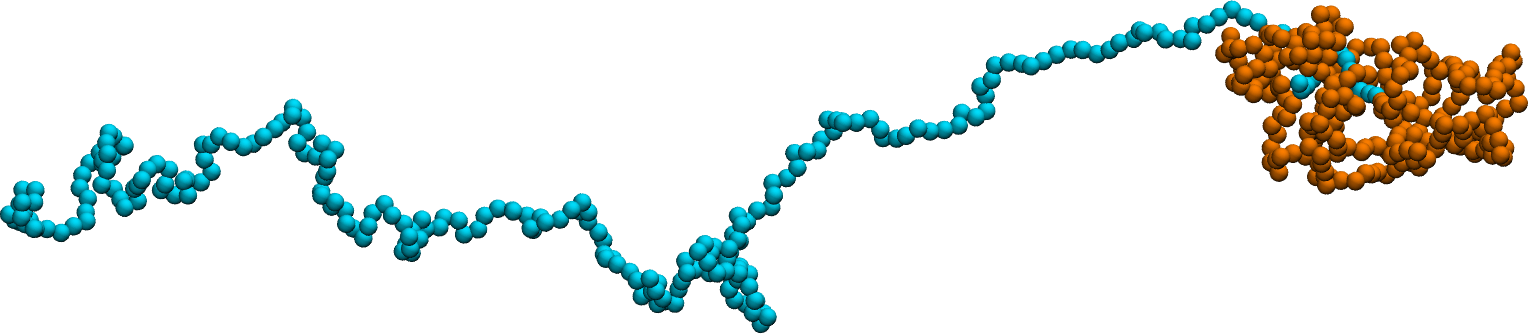} \\
(b) \\
\includegraphics[width=1.\columnwidth]{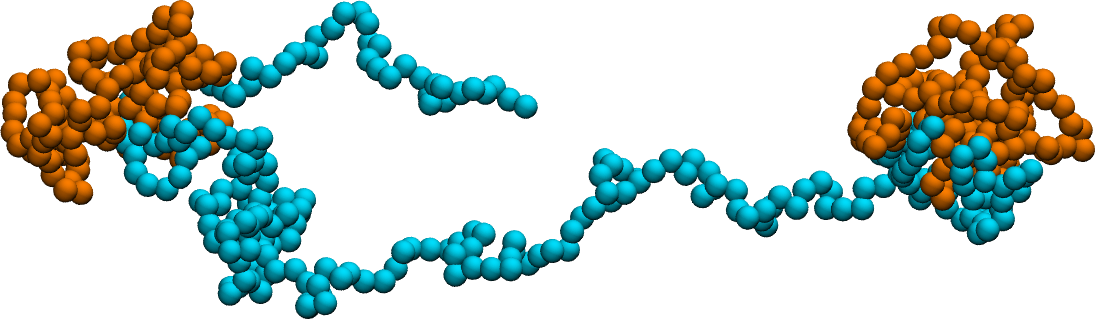} \\
(c) \\
\includegraphics[width=1.\columnwidth]{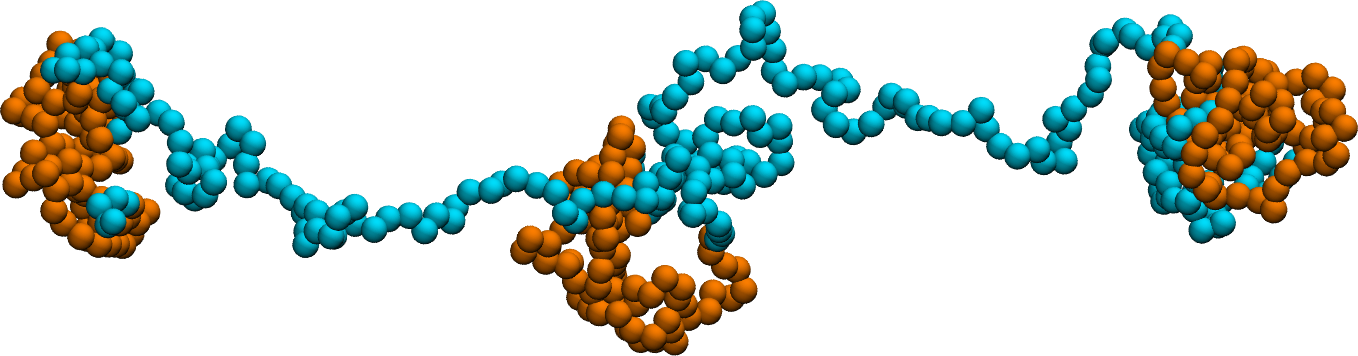} \\
(d)
\end{tabular}
\caption{Snapshots of steady state polymer configurations with $N=500$, $p=0.5$, and $\mathrm{Pe}=10$ for different arrangements of active (orange) and passive (blue) monomers along the chain. (a) random distribution of active monomers, (b) one contiguous active section, (c) two contiguous non-overlapping active sections and (d) three contiguous non-overlapping active sections.}
\label{fig:configurations}
\end{figure}

We model the polymer as a fully flexible, self-avoiding bead-spring linear chain consisting of $N$ monomers, suspended in a bulk fluid in three dimensions. The self-avoidance between any pair of monomers is implemented via a truncated {and shifted} Lennard-Jones (LJ) potential:
\begin{equation}
V_{\text{LJ}}(r) =
\begin{cases}
  4\epsilon \left[ \left(\frac{\sigma}{r} \right)^{12}- \left(\frac{\sigma}{r}\right)^{6}+\frac{1}{4}\right] & \text{for~}  r < 2^{1/6}\sigma \\
  0 & \text{for~} r \geq 2^{1/6}\sigma
\end{cases}
\end{equation}
where {$\sigma=1$ is the diameter of the monomer and is taken as the unit of length,} $\epsilon=10 k_B T$ sets the interaction energy and $r=|\vec{r}_i - \vec{r}_j|$ is the Euclidean distance between the monomers $i$ and $j$ positioned at $\vec{r}_i$ and $\vec{r}_j$, respectively. {We take the thermal energy as the unit of energy, $k_B T=1$.} In addition, the Finitely Extensible Nonlinear Elastic (FENE) bonding potential~\cite{kremer1990dynamics}
\begin{equation}
	V_\mathrm{FENE}(r) = -\frac{K r_0^2}{2} \ln \left[ 1 - \left( \frac{r}{r_\mathrm{0}} \right)^2 \right]
\end{equation}
acts between any pair of consecutive monomers along the polymer backbone. We set $K=30\epsilon/\sigma^2$ {=$300 k_\mathrm{B}T/\sigma^2$} and $r_0=1.5\sigma$ to {avoid strand crossings}. {The activity is introduced as a tangential self-propulsion\cite{bianco2018globulelike}: on a given monomer $i$ at position $\vec{r}_{i}$, acts an active force 
\begin{equation}
\vec{f}^{\mathrm{a}}_{i} = f^{\mathrm{a}} \hat{t}_i    
\end{equation}
where {$\hat{t}_i = (\vec{r}_{i+1} - \vec{r}_{i-1})/|\vec{r}_{i+1} - \vec{r}_{i-1}|$} is the normalized tangent vector. The magnitude of the active force on each active monomer $f^{\mathrm{a}}$ is constant; the strength of the activity is controlled, as usual, by varying a dimensionless parameter called the P\'eclet number $\mathrm{Pe}=|{f_\mathrm{a}}|\sigma/k_\mathrm{B}T$. We consider here the case where} only a fraction $p$ of the monomers is {active, while the} remaining monomers are passive. The active monomers along the chain {are organized in contiguous sections, or blocks, characterized by their starting contour position. We will mainly focus on the case where only one active block is present; for the sake of highlighting the special nature of this arrangement, we will also briefly consider polymer chains with two and three sections, as well as the case of randomly arranged, un-clustered active sites. We report, in Fig.~\ref{fig:configurations} a few snapshots, exemplifying the different types of the active sites' organization considered in this work. We remark that the introduction of the tangential self-propulsion breaks the symmetry of the polymer and allows to define a head and a tail of the chain. Naturally, one would follow the index of the monomers and call ``head'' the first monomer and ``tail'' the last. By construction, the self-propulsion forces point towards the tail of the chain, as each one points towards the monomer with a higher index ($i+1$). However, without loss of generality, we will switch places between "head" and "tail" in the rest of the manuscript.} 

\subsection{Simulation details}
\label{sec:sim}
{We study isolated, partially active polymers in bulk by means of Langevin Dynamics simulations using the open source code LAMMPS\cite{thompson2022lammps}, with in-house modifications to implement the tangential activity; we neglect hydrodynamic interactions. Bulk conditions are implemented with periodic boundary conditions: in order to} exclude the effect of self-interaction of the chain across the periodic boundary, the box side is {chosen so that the each size is slightly larger than the polymer's contour length.}
{The equations of motion are integrated using the velocity Verlet algorithm, with elementary time step $\Delta t = 10^{-3}$. We set $m=1$, $\sigma=1$, $k_B T = 1$ as the units of mass, length, and energy, respectively; the unit of time is $\tau= \sqrt{m \sigma^2/ k_B T} = 1$. In order to ensure the overdamped regime\cite{fazelzadeh2022effects}, we set the friction coefficient $\gamma=20 \tau^{-1}$.}


{We study polymers of different length} {$100 \leq N \leq 750$}, {different percentage of active sites} {$0.1 \leq p \leq 0.5$} and {different valued of the activity } {$0.1 \leq \mathrm{Pe} \leq 10$}, {although we will mostly focus on the high activity case $\mathrm{Pe}=10$}. For a given set of parameters, {after reaching a steady state, production runs have been performed for, at least, }$7.5 \cdot 10^8\tau$. Polymer conformations have been sampled at a rate of $10^7 \Delta t= 10^4 \tau$, that is larger than {the decorrelation time of the end-to-end vector}. Furthermore, if not specifically mentioned, {$M=$25 independent trajectories were simulated, in order to improve the statistical significance.}

\subsection{Metric and dynamical properties}
\label{sec:obs}
In this section, we will introduce the different metric and dynamical properties that will be considered in our study.\\
The {metric properties} of a polymer describe its size and shape; as standard practice, we compute the eigenvalues and eigenvectors of the gyration tensor
\begin{equation}
G_{\alpha \beta} = \frac{1}{N}\sum_{i=1}^{N}{(\textbf{r}_{i,\alpha} - \textbf{r}_{cm,\alpha})(\textbf{r}_{i,\beta} - \textbf{r}_{cm,\beta})}    
\end{equation}
where $\textbf{r}_i$ is the coordinate of the $i$-th monomer, $\textbf{r}_{cm}$ is the coordinate of the center of mass $\textbf{r}_{cm} = 1/N \sum_{i=1}^{N}{\textbf{r}_i}$ and $\alpha$ and $\beta$ stand for the three Cartesian coordinates. We compute the three eigenvalues $\lambda_1$, $\lambda_2$, $\lambda_3$ ($\lambda_1 \geq \lambda_2 \geq \lambda_3$) for each polymer conformation in steady state. From these three values, one can compute the gyration radius as $R_g^2 = \lambda_1 + \lambda_2 + \lambda_3$; $R_g$ gives an estimate of the spatial extension of the polymer. Further, the relative shape anisotropy $\delta^*$ and the prolateness $S^*$~\cite{narros2013effects} can be computed. 
The relative shape anisotropy is given by
\begin{equation}\label{eq:anisotropy}
\delta^*=1-3 \left\langle \frac{I_2}{I_1 ^ 2} \right\rangle
\end{equation} 
while the prolateness is given by 
\begin{equation}\label{eq:prolateness}
S^* = \left\langle \frac{(3\lambda_1 - I_1)(3\lambda_2 - I_1)(3\lambda_3 - I_1)}{I_1 ^3} \right\rangle
\end{equation} 
where in both cases the average is done over time and over the ensemble of the independent realizations. Further, $I_1$ and $I_2$ are defined as
\begin{align} 
I_1&=\lambda_1+\lambda_2+\lambda_3 \cr
I_2&=\lambda_1\lambda_2+\lambda_2\lambda_3+\lambda_3\lambda_1 \nonumber
\end{align}
The prolateness is zero for spherical objects, assumes negative values for oblate (disk-like) shapes and positive values for prolate shapes. The shape anisotropy vanishes for high symmetric configurations and is positive otherwise. Furthermore, we will consider the correlation of the tangent vector along the chain
\begin{equation}
    C(s) = \langle \hat{t}(s_0+s) \cdot \hat{t}(s_0) \rangle
\end{equation}
where $t(s)$ is the tangent vector at the contour position $s$ and $\hat{t}(s)=t(s)/|t(s)|$ is the corresponding unit vector. The average is done on the initial contour position $s_0$, on time (in steady state) and on the independent realizations. We will compute this function in the passive part of the chain, as an indirect estimator of the shape and of the (effective) semi-flexibility of the chain.\\Concerning the dynamical properties, we will characterize the mobility of the active polymers via the mean square displacement (MSD) of the polymer center of mass, located at $\vec{r}_{cm}$, defined as
\begin{equation}
    \Delta R^2(t) = \langle (\vec{r}_{cm}(t_0+t) - \vec{r}_{cm}(t_0))^2 \rangle
\end{equation}
We will also consider the MSD of the central monomer of the chain, defined as 
\begin{equation}
    \Delta R^{2 \star}(t) = \langle (\vec{r}_{N/2}(t_0+t) - \vec{r}_{N/2}(t_0))^2\rangle
\end{equation}
Further, we will compute the autocorrelation function of the active section
\begin{equation}
    \chi (t) = \left\langle \frac{R^a_e (t+t_0) \cdot R^a_e (t_0)}{|R^a_e (t+t_0)||R^a_e (t_0)|} \right\rangle
\end{equation}
where $R^a_e$ is the end-to-end vector of the active block, i.e. the vector that connects the first to the last active monomer. In the case of $\Delta R^2(t)$, $\Delta R^{2\star}(t)$ and $\chi(t)$, the average is taken over the initial time $t_0$ and over the independent realizations. Finally, we will also compute the non-Gaussian parameter
\begin{equation}
	\alpha_2(t) = \frac{3}{5} \frac{\langle \Delta r^4(t) \rangle}{\langle \Delta r^2(t) \rangle^2} - 1,
\end{equation}
The average is, in this case, taken over the independent configurations; we then report the mean value over time in the steady state $\alpha_2 = \langle \alpha_2(t) \rangle$. Any non-zero value of $\alpha_2$ highlights a deviation of the displacement distribution from a Gaussian. 

\section{Results}
\subsection{Dynamical heterogeneity of a population of isolated partially active polymers}
\label{sec:nongauss}
{First, we show that active blocks introduce dynamical heterogeneity at the population level. We consider four different settings: a) the active monomers are distributed at random along the chain, b) the active monomers are arranged in one active block, c) the active monomers are arranged in two, non-overlapping active blocks, d) the active monomers are arranged in three, non-overlapping active blocks. For each case, we consider different values of the fraction of the active monomers $p \in [0.1, 0.5]$. We take care that, in all cases, the effective number of active monomers is the same, so that the overall activity is equal; in particular, in case a), a monomer can't be picked twice in the random selection process and in cases c-d) there is no overlap between the active blocks. We further consider three values of the P\'eclet number $\mathrm{Pe}=$0.1, 1, 10. For a given set of values of $p$ and $\mathrm{Pe}$, we simulate a population of $M^*=$100 polymer chains of fixed length $N=$500 monomers. Within case a), different polymers have a different random arrangement; within cases b-d), the contour position of the first monomer of the active block(s) is chosen, for each of the $M^*$ polymers, at random.}
\begin{figure}[!hb]
	\centering
	\includegraphics[width=1\columnwidth]{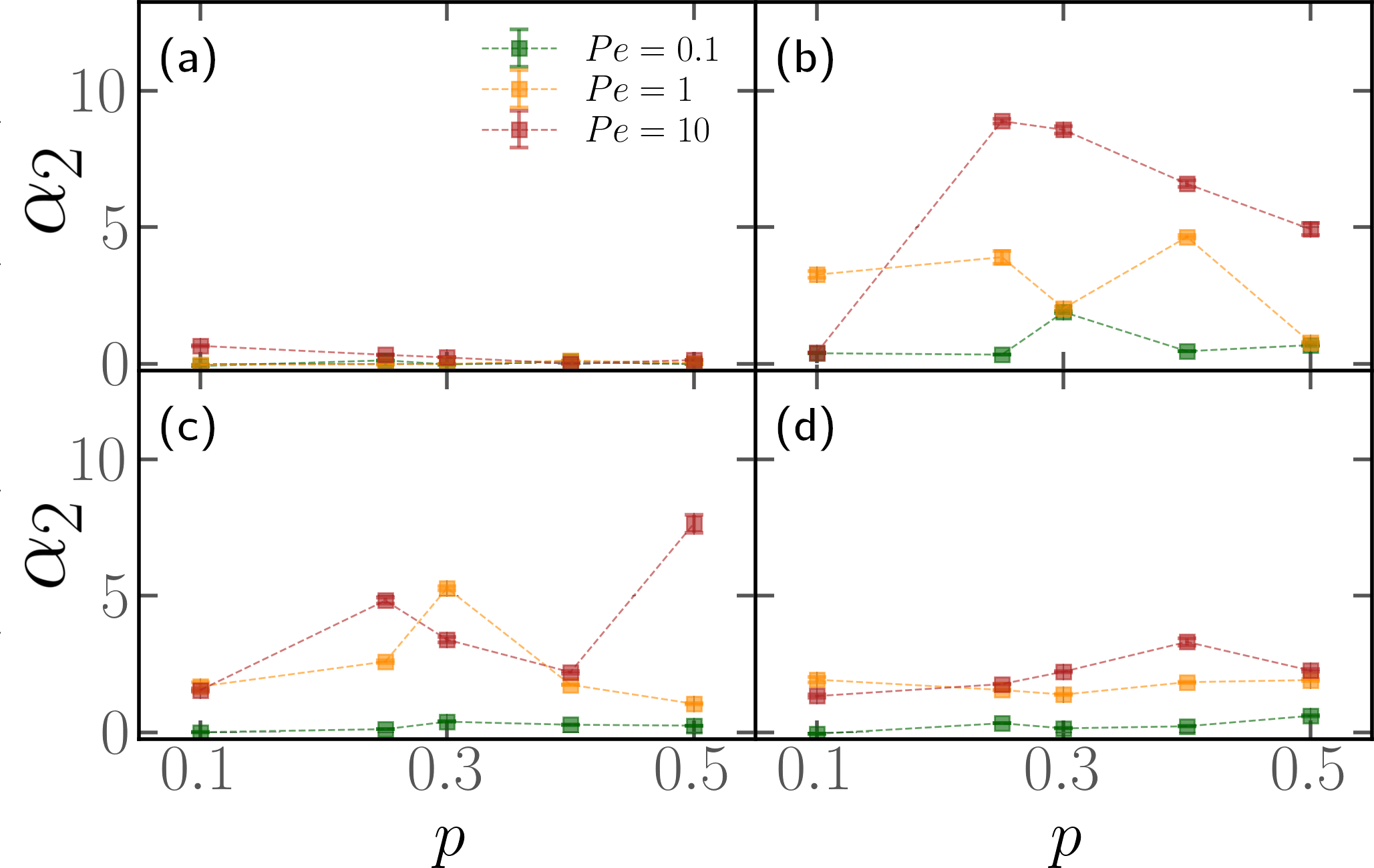}
	\caption{Time-averaged Non-Gaussian parameter $\alpha_2$ as a function of the percentage of active monomers $p$ for different values of $\mathrm{Pe}$, and \textbf{(a)} random distribution of active monomers, \textbf{(b)} one active block, \textbf{(c)} two non-overlapping active blocks \textbf{(d)} three non-overlapping active blocks. In all cases, we considered populations of $M^*=$100 polymers, made of $N =$500 monomers. Dotted lines are guides to the eye.}
	\label{fig:NGP}
\end{figure}
{The results are reported in Fig.~\ref{fig:NGP}, where we show the non-Gaussian parameter $\alpha_2$, defined in Sec.~\ref{sec:obs} as a function of the fraction of active monomers $p$ for different values of $\mathrm{Pe}$; the four panels refer to the four cases a)-d). We observe the emergence of a non-zero dynamical heterogeneity for cases b-d); conversely, a) shows negligible heterogeneity, in a population of the same size, for all values of $p$ or $\mathrm{Pe}$ considered. This suggests that arranging the active monomers at random does not lead to qualitative difference in the dynamics of the different polymers. The same is not true for the b-d) cases, where the non-Gaussian parameter $\alpha_2$ is sensibly larger than one when the activity of the individual monomers is large enough (i.e. $\mathrm{Pe}\gtrsim$1).}\\
{We also notice that the difference in the dynamics is reflected in the polymer conformations. One can notice} in Fig.~\ref{fig:configurations}, {that the conformations in the cases b-d) are markedly different from the case a). In particular, in Fig.~\ref{fig:configurations}a the chain exhibits a coil-like polymer conformation; on the contrary, when the active monomers are grouped in contiguous sections (Fig.~\ref{fig:configurations}b-d) the active regions substantially shrink, while the passive regions elongate. As we will see in detail in the case of polymers with a single active block, this depends on the size of the section itself and on the position of its first monomer, leading to conformational heterogeneity. Both scenarios are different from the fully active case, that is characterized by globule-like conformations\cite{bianco2018globulelike, foglino2019non}}.\\
From a dynamical perspective, the coil-like configuration in case a) can be understood as follows. At small values of $p$, placing the monomers at random along the chain does not lead to clustering of the active sites, for an overwhelming large number of realizations. In this regime, the tangential activity will still influence the local conformation of the polymer; however, the active forces, acting on monomers that are separated along the contour, will not be correlated and will not influence the conformation on the scale of the whole chain. Upon increasing the value of $p$, more and more small clusters will appear, until the fully active case is recovered; indeed, the conformation shown in Fig.~\ref{fig:configurations}, at $p=$0.5, already resembles the globule-like state\cite{bianco2018globulelike}. The same argument can be further carried to the dynamics: as long as the effect of the self-propulsion remains local, different random arrangements will not lead to different dynamical behaviour.


{Finally, notice that $\alpha_2$ becomes smaller as the number of active blocks increases. The random arrangement can be seen as a limiting case and it has, indeed, negligible non-Gaussianity. This can be understood in a statistical sense; the detailed calculations are reported in the Supplemental Material. 
In brief, the set of possible arrangements is maximal for $N \cdot p$ clusters of size 1, i.e. the random case. Contiguous section are possible as they represent a subset of all the possible random arrangements, but are overwhelmingly rare; even though their dynamical and conformational properties are markedly different from the ones of the average random arrangement, their importance is negligible.}
{From this point on, we will focus on the case of a single active block at high values of $\mathrm{Pe}$ ($\mathrm{Pe}$=10); data concerning different values of $\mathrm{Pe}$ are reported in the Supplemental Material. } {We make this choice essentially for two reasons. First, the set of possible arrangements for a single active block is the smallest and one can introduce a parameter that greatly helps rationalizing the rich dynamical scenario. Second, as shown, the effects of the heterogeneous activity are, in this case, more evident.} 

\subsection{Contour position of the active block determines shape and size of the chain}
\label{sec:rgx}
{In order to quantify the position of the active block along the contour,} {we introduce the parameter $x=N_\text{p}/N$, $N_\text{p}$ being the number of passive monomers between the head of the polymer (see Sec.~\ref{sec:model}) and the first monomer of the active section; alternatively, it is the minimum contour distance between the head of the polymer and the edge of the active section. The minimum value of $x$ is $x=1/N$ as, for the end monomers, a tangent vector can't be defined and, thus, they are always passive; the maximum value is $x=1-p-1/N \approx 1-p$ for large $N$. In general, if $x\approx 0$, the active section is close to the head of the polymer; upon increasing $x$, the active section shifts towards the tail.}\\
\begin{figure}[!h]
	\centering
	\includegraphics[width=0.8\columnwidth]{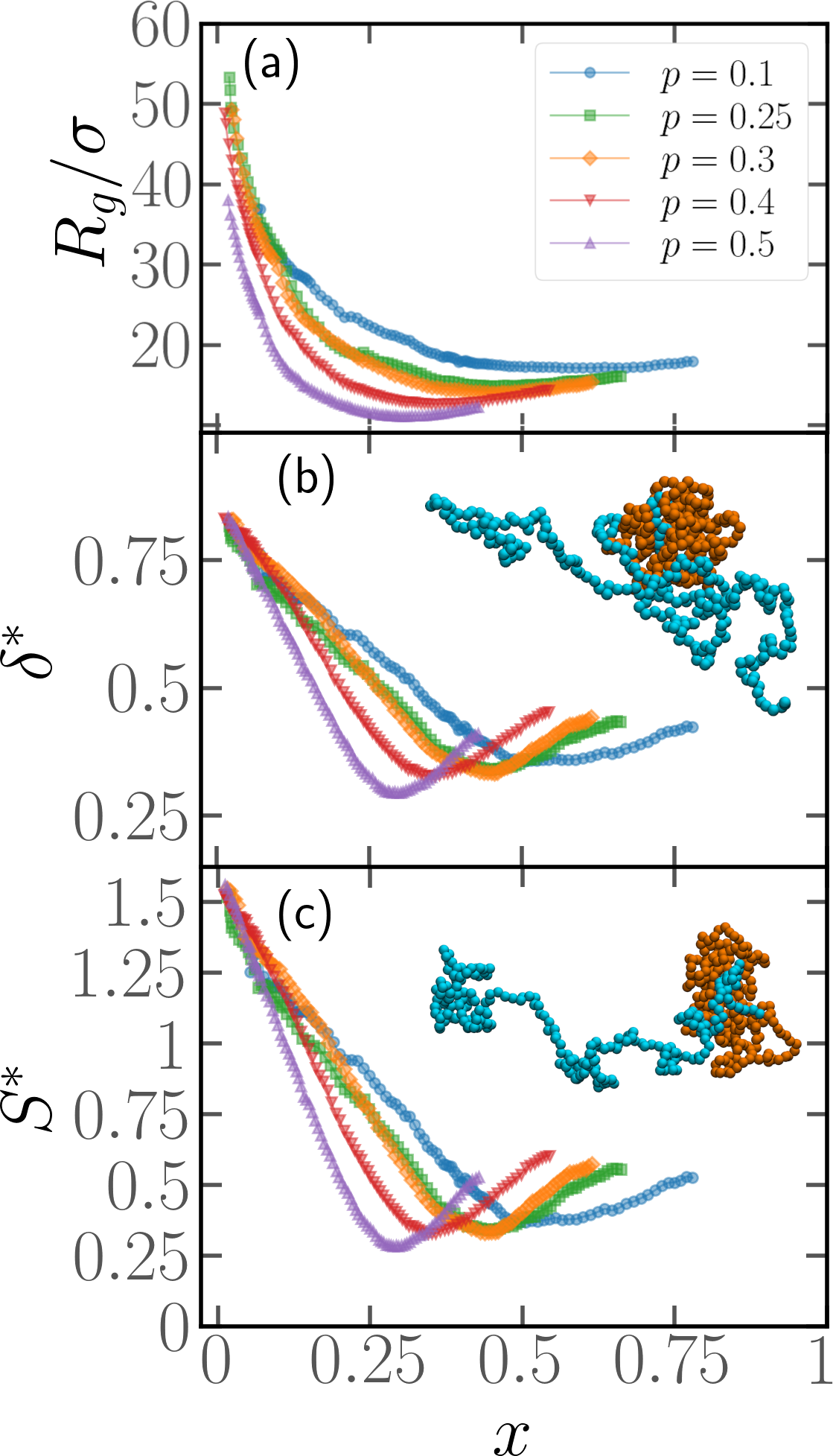}
	\caption{Gyration Radius (a), Asphericity (b), and Prolateness (c) as a function of the ratio of the number of passive monomers at the end of the chain to the total number of monomers, $x$, for $Pe=10$, $N = 500$ and for different values of the percentage of active monomers $p$. Snapshots in panels (b),(c) show typical polymer conformations for fixed $p=$0.5 and (b) $x=(1-p)/2$, (c) $x=1-p$ .}
	\label{fig:geo_q_0}
\end{figure}
We present, in Fig.~\ref{fig:geo_q_0}, how the metric properties, i.e. the gyration radius $R_g$, the shape anisotropy $\delta^*$ and the prolateness $S^*$, depend on the parameter $x$. We consider here polymer chains of length $N=$500, $\mathrm{Pe}=$10, and different values of $p$. We expect a polymer of size $N=$500 to be in the scaling regime (see Sec.~\ref{sec:scalingRg}). The results reported are obtained from the same set of simulations described in Sec.~\ref{sec:nongauss}, i.e. a population of $M^*$ polymers with one active block, whose position has been chosen at random along the chain. This implies that very few chains over the total will be characterized by the same value of $x$; in order to better visualize the trends, the data in Fig.~\ref{fig:geo_q_0} has been smoothed by means of a convolution filter. 
All the curves show non-monotonic trends, regardless of the smoothing. \\We start from Fig.~\ref{fig:geo_q_0}a, where the gyration radius as a function of $x$ is reported. We observe that $R_g$ decreases from its maximum value at $x=1/N$ upon increasing $x$; the minimum value is reached at $x = x_{\mathrm{min}} \approx (1-p)/2$, and at higher values of $x$, $R_g$ increases again, becoming non-monotonic. Similar trends are visible in Fig.~\ref{fig:geo_q_0}b,c for the asphericity and the prolateness, respectively; we notice that the minimum is more pronounced. We also notice that the prolateness is always positive; the polymers, upon increasing $x$, transition from a very elongated shape to a more spherical one and become elongated again above $x_{\mathrm{min}}$. At variance with $R_g$, the maximum value at $x=1/N$ in Fig.~\ref{fig:geo_q_0}b,c appears almost independent on $p$; this indicates that the shape of the active polymers remains the same in the limit $x \to 1/N$, within the range of values $0.1 \leq p \leq 0.5$ considered in this work. As a further comment, notice that the active block is, for the values of $p$ considered, long enough such that the active section attains a globule-like configuration, due to buckling-like instability\cite{bianco2018globulelike, foglino2019non} and it is thus pretty compact and spherical. Thus, the overall prolateness of the chain is essentially due to the passive section.\\Notice that, for $x_{\text{min}} = (1-p)/2$, the center of the active block is located exactly halfway along the contour of the polymer: the non-monotonicity in Fig.~\ref{fig:geo_q_0} indicates that any shift of the location of the active block from the middle of the chain leads to the elongation of the entire polymer chain. However, the curves are not symmetric about $x_{\text{min}}$, a further consequence of the broken symmetry introduced by the tangential activity; thus it is the position of the active block with respect to the head of the chains that determines the overall conformation.\\To exemplify, the snapshot reported in Fig.~\ref{fig:configurations}b refers to a polymer with $x=1/N$; in the insets of Fig.~\ref{fig:geo_q_0} we report snapshots of polymers with $x=(1-p)/2$ (inset of Fig.~\ref{fig:geo_q_0}b) and $x=1-p$ (inset of Fig.~\ref{fig:geo_q_0}c). The simplest case is $x=1/N$, where the active block pulls the passive section, that elongates. For $x=1-p$ the active block also ends up pulling the passive section, which rationalizes the observed elongation; however, it does so in a much less efficient way, resulting in a partial folding. Further, for $x=(1-p)/2$, the active block is positioned between two, equally sized blocks; both are pulled around by the active section, which leads to complete folding and a more spherical shape. Notably, some of these features still persist if the polymer chain has two or three active blocks (see Supplemental Material). 


\subsection{Effective persistence of the passive section}
\label{sec:passive}

In order to complement the analysis carried out so far, we aim to characterize the typical conformations of the passive section. We will focus here on the longest passive section, of length $N_{p}^{l}$ and we will consider the tangent-tangent correlation function $C(s)$, defined in Sec.~\ref{sec:obs}, as a function of the contour distance $s/N$.

\begin{figure}[!h]
	\centering
	\includegraphics[width=0.8\columnwidth]{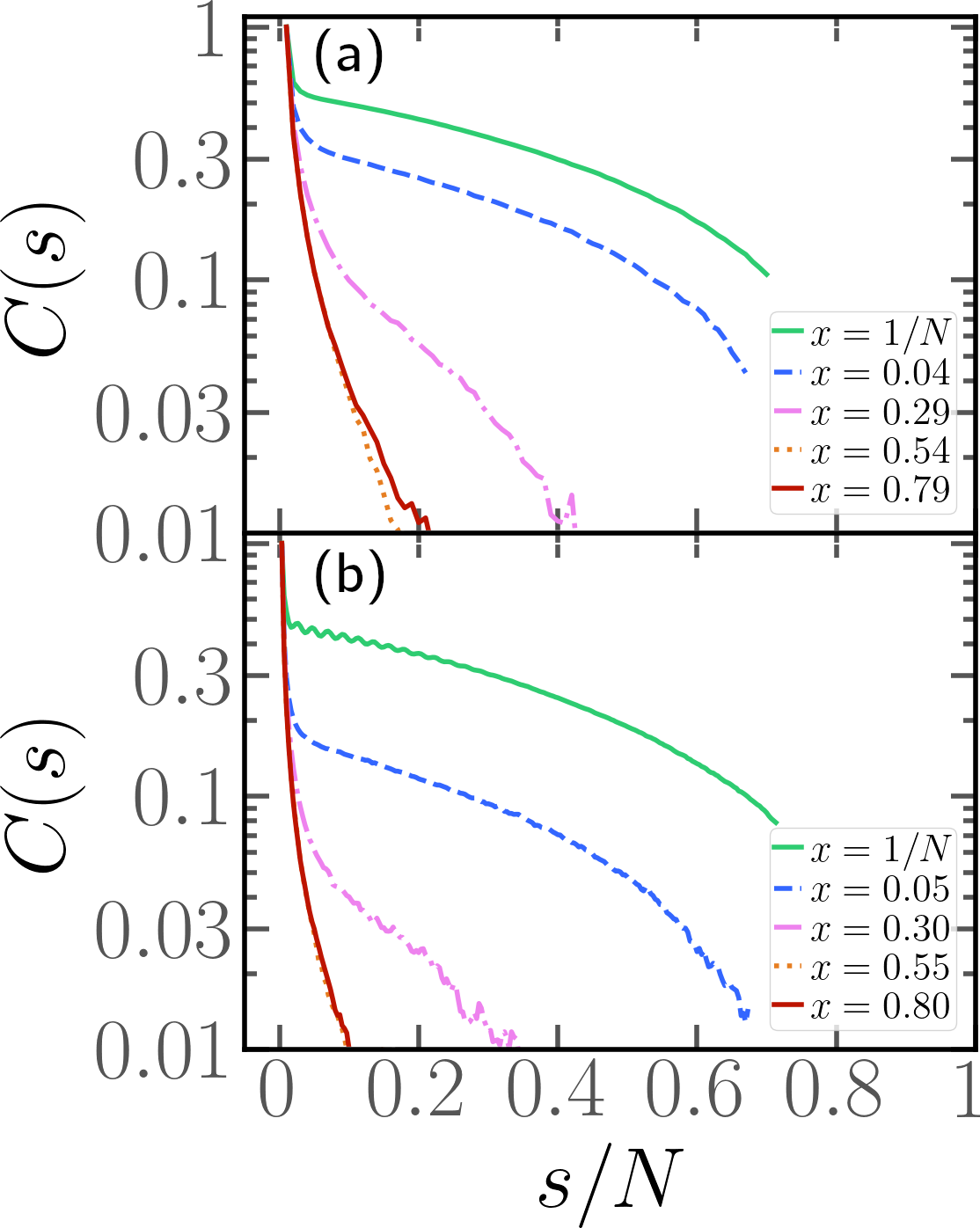}
	\caption{Tangent-tangent correlation function as a function of the contour distance $s$ within the longest passive section for $p = 0.2$, $\mathrm{Pe} =$10, and (a) $N = 100$, (b) $N = 300$.}
	\label{fig:pl}
\end{figure}

Looking at the conformations, e.g. Figs.~\ref{fig:configurations},~\ref{fig:geo_q_0}, the passive section appears always stretched and elongated, albeit to different degrees at different values of $x$. One way to recast this property is to introduce an effective rigidity. To provide a quantitative description of this phenomenon, we indeed look at the tangent-tangent correlation function within the largest passive section. The results are presented in Fig.~\ref{fig:pl} for $N=$100, 300, $p=$ 0.2, and different values of $x$. We observe that $C(s)$ depends strongly on $x$: when the parameter $x$ is small, the tangent vectors remain correlated throughout the whole passive section. However, upon increasing $x$, the correlation decays more rapidly: it is easy to see that, at high enough values of $x$, the correlation length becomes negligible. This analysis thus allows us to assess how the positioning of the active sections along the chain influences the effective persistence length of the passive section: the closer the active cluster is to the head of the chain, the higher the effective persistence length. We therefore expect that upon increasing the degree of polymerization $N$, chains characterized by a small value of $x$ will grow with a large scaling exponent, similar to rigid rods. On the contrary, chains characterized by a large value of $x$ will grow with a scaling exponent, comparable to the passive one for self-avoiding chains. 


\subsection{Scaling properties for polymer chains with one active block}
\label{sec:scalingRg}
We now assess how the contour position of the active block affects the scaling properties of the gyration radius, i.e. how $R_g$ depends on the degree of polymerization $N$.
\begin{figure}[!h]
	\centering
	\includegraphics[width=0.8\columnwidth]{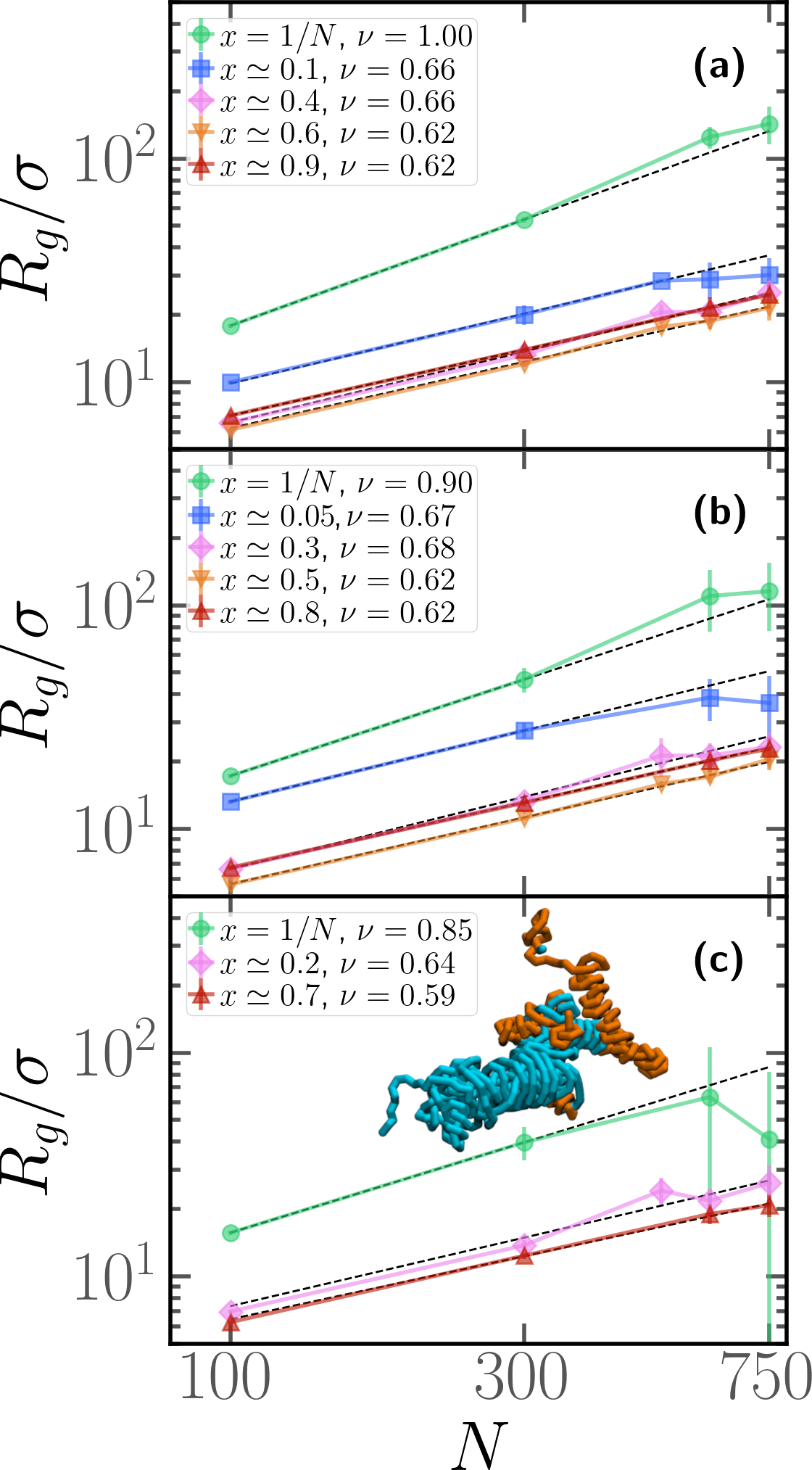}
	\caption{Gyration radius as a function of the degree of polymerization $N$ for different values of $x$ and (a) $p=0.1$, (b) $p=0.2$, (c) $p=0.3$. Full lines are guide to the eye, dashed lines are power-law fits $R_g = \alpha N^{\nu}$; the result of the fit is reported in the legend. Snapshot in the inset of panel (c) refers to a severely entangled configuration observed at $N=$600, $\mathrm{Pe}=$10, $x=1/N$, $p=0.3$. }
	\label{fig:rg}
\end{figure}
We report the results in Fig.~\ref{fig:rg}; in each panel, we report the gyration radius as a function of the degree of polymerization $N$ for different values of $x$. In this case, we performed, for each value of $x$, $M=$25 independent realizations, i.e. $M$ independent copies with the same arrangement of active monomers. Notice that the value of $x$, reported in the legend, is only approximately similar for polymers with different $N$. In our simulations, we fixed for convenience the contour positions, in an arbitrary fashion; due to the presence of the ``head'' bead, which is always passive, there is a $1/N$ contribution to the value of $x$ that was disregarded. In the different panels, we report results for different values of the fraction of active monomers $p$. We observe that the gyration radius follows, for all values of $x$ and $p$, a power law $R_g \sim N^{\nu}$; however, clearly, upon changing $x$, the scaling exponent $\nu$ changes drastically. In all three panels, we observe that if $x=1/N$ the scaling exponent is quite high, compared to the passive exponent $\nu=$0.588 of self-avoiding polymers; this is in agreement with the extremely high value of the prolateness, observed in Fig.~\ref{fig:geo_q_0} at fixed $N$ and with the effective persistence of the passive section, reported in Sec.~\ref{sec:passive}. Notice, however, that the exponent slightly decreases upon increasing $p$. Indeed, upon increasing $p$, the contribution of the active section, which attains a globule-like conformation, becomes more important.\\Upon shifting the active block along the contour to larger values of $x$, the scaling exponent becomes $\nu \simeq 0.62$, slightly larger than the passive scaling exponent $\nu=$0.588. As observed in Sec.~\ref{sec:rgx}, the passive sections must be rather elongated, in order to justify the overall prolateness of the chain; this may be sufficient, overall, to account for the scaling exponent observed. In agreement with the data reported in Fig.~\ref{fig:geo_q_0} at fixed $N$, the non-monotonicity of $R_g$ is visible also in panels a) and b) of Fig.~\ref{fig:rg}, and curves at fixed $x\simeq 0.5$ display the lowest value of $R_g$. 


\begin{figure*}[!h]
	\centering
		\includegraphics[width=0.75\textwidth]{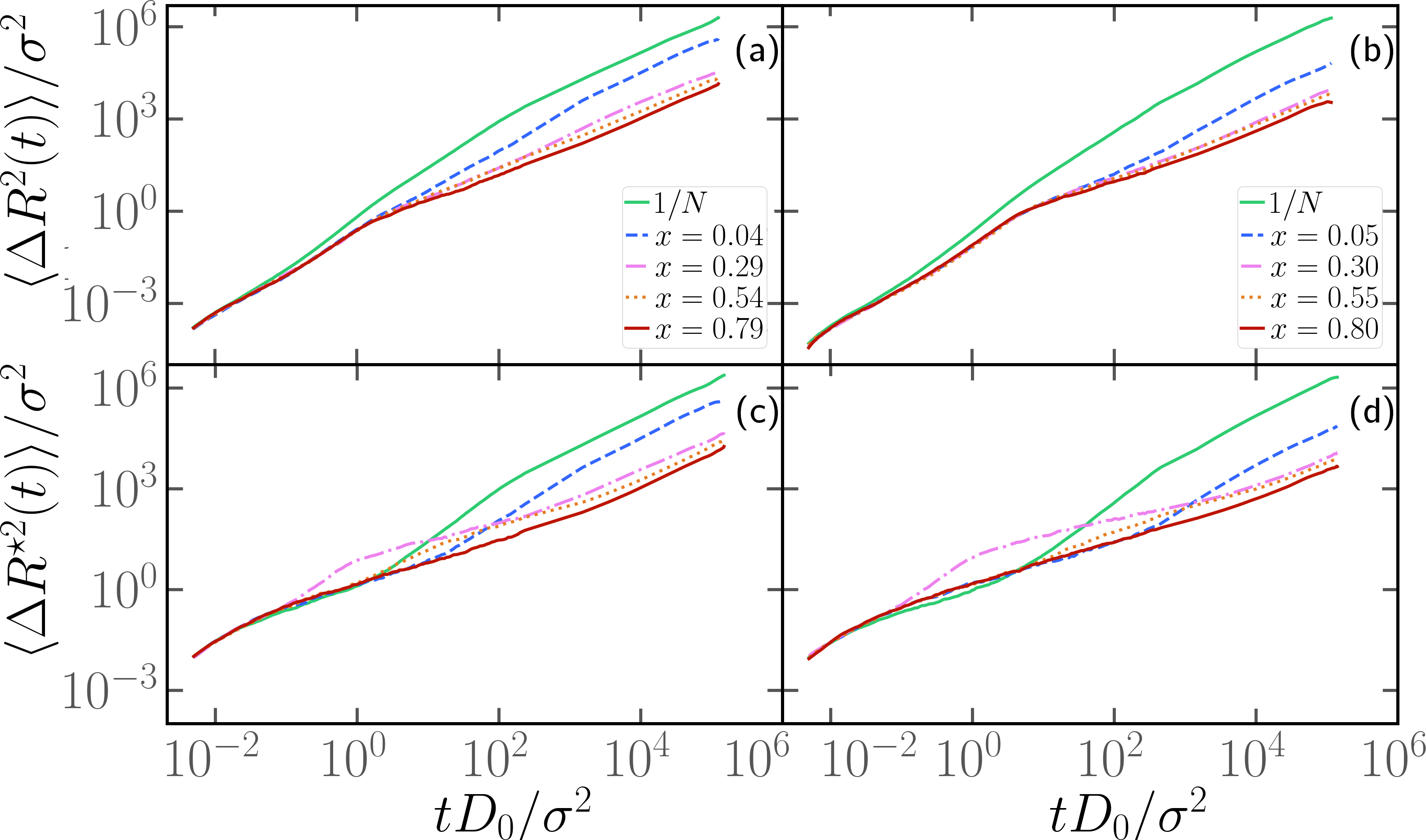}\hfil
		\caption{(a-b) Log-log plot of the mean square displacement of the center of mass of partially active polymers as a function of time for different values of $x$ and (a) $N = 100$, (b) $N = 300$. (c-d) Mean square displacement of the central monomer as a function of time for different values of $x$ and (c) $N = 100$, (d) $N = 300$. In all panels, $p=0.2$ and $\mathrm{Pe}=$10 are fixed.}
	\label{fig:msd_long}
\end{figure*}


{Finally, we highlight a significant deviation of the gyration radius from the global trend for $x=1/N$, with $N\geq 600$ in Fig.~\ref{fig:rg}c ($p=$0.2, 0.3); these data, that deviate from the power law, are accompanied by large error bars. This phenomenon is due to self-entanglements: if the polymers are long enough and the active block is placed very close to the head of the chain, we obtain a spontaneous and abundant knot formation. A fraction of the polymers display numerous knots within the same chain but remain very elongated, others get tangled and end up in a very compact state (see inset of Fig.~\ref{fig:rg}c); the size of the error bars reflects this heterogeneity. The same phenomenology was also observed in sufficiently long isolated active rings\cite{locatelli2021activity} and was also attributed to entanglement.} {This particularly intriguing phenomenon is not the focus of this article and will be addressed in more detail in a future work. We have not considered the data points corresponding to these situations ($x=1/N$, $N\geq600$) for the fit presented in Fig.\ref{fig:rg}.}

\subsection{Dynamics of polymer chains with one active section}
\label{sec:msd}

We now turn to the discussion of the dynamics of the partially active polymer. As seen in the fully active case, the introduction of the tangential activity ties conformation and dynamics together. We first discuss the dynamics by looking at the Mean Square displacement of the center of mass (see Sec.~\ref{sec:obs}) as a function of the rescaled time $t D_0/\sigma^2$, reported in panels a,b of Fig.~\ref{fig:msd_long} for $N=$100 and $N=$300, respectively; $D_0$ refers to the diffusion coefficient of a single passive monomer. Also, in this case, as in Sec.~\ref{sec:scalingRg}, we consider $M=$25 independent realizations, with the same arrangement of active monomers. We have fixed $p=$0.2 and $\mathrm{Pe}=$10 in Fig.~\ref{fig:msd_long}; results for different values of $p$ are reported in the Supplemental Material. In both panels of Fig.~\ref{fig:msd_long} we observe that the extreme case $x = 1/N$ exhibits the largest MSD. All the curves overlap up to $t D_0/\sigma^2 \sim 1$ showcasing a common super-diffusive regime; then, for $t D_0/\sigma^2 > 1$, polymers characterized by a large value of $x$ slow down and become diffusive. Notice that the MSD shows a sub-diffusive regime, that becomes more evident upon increasing $N$. This signals that, at intermediate time scales, the polymers tend to have relatively long ``tumbling'' periods, when they rotate around their center of mass. Instead, polymers with a relatively small value of $x$, up to extreme value $x=1/N$ remain super-diffusive for a much longer time; in contrast with the large $x$ case, changing the position of the active block along the contour by a few monomers changes the diffusion coefficient significantly (see also Sec.~\ref{sec:Dx}).\\We further look at the monomeric MSD; we focus on the MSD of the central monomer of the chain (see \ref{sec:obs}) for convenience. In fact, the average over all monomers, also known as $g_1(t)$ in the literature, would depend critically on the value of $p$ and would not really highlight the dynamics of either region, active or passive. We report the results of $\langle \Delta R^{\star 2}(t) \rangle/\sigma^2$ as a function of the rescaled time $t D_0/\sigma^2$, in Fig.~\ref{fig:msd_long}c,d. We observe that the MSD of the central monomer depends strongly on $x$. We showed in Fig.~\ref{fig:msd_long}a,b that the position of the active block along the contour modifies the overall dynamics. On top of this, depending on the value of $x$, the central monomer can be either passive or active; in particular, referring to the values of $x$ reported in Fig.~\ref{fig:msd_long}c,d, the central monomer is active only for the value $x\simeq$ 0.29-0.30. At the lowest value of $x$, $x=1/N$, $\langle \Delta R^{\star 2}(t) \rangle/\sigma^2$ shows, after an initial sub-diffusive regime, a strong super-diffusion followed again by normal diffusion, signaling the transition to the center of mass dynamics. The short-time sub-diffusive regime lasts longer upon increasing $x$, if the central monomer remains passive. However, when the central monomer becomes active ($x\simeq$ 0.29-0.30), a more complex behaviour emerges: after the short-time sub-diffusive regime, a brief super-diffusive regime appears, followed by a new, rather long-lasting sub-diffusive regime, roughly three decades in time. This sub-diffusive regime is related to the globule-like regime and overlaps with the sub-diffusive regime observed for the center of mass dynamics (Fig.~\ref{fig:msd_long}a,b). Upon increasing $x$ further, the super-diffusion shifts to later times and the intermediate sub-diffusive regime shrinks; at sufficiently high values of $x$, they are both suppressed and a passive-like behaviour is recovered. Interestingly, this latter feature distinguishes $\langle \Delta R^{\star 2}(t) \rangle/\sigma^2$ from the MSD of the center of mass, which shows an anomalous sub-diffusive behaviour for all values of $x$ considered.


\subsection{Contour position and size of the active block determine the mobility of the chain}
\label{sec:Dx}

\begin{figure}[!h]
	\centering
	\includegraphics[width=0.8\columnwidth]{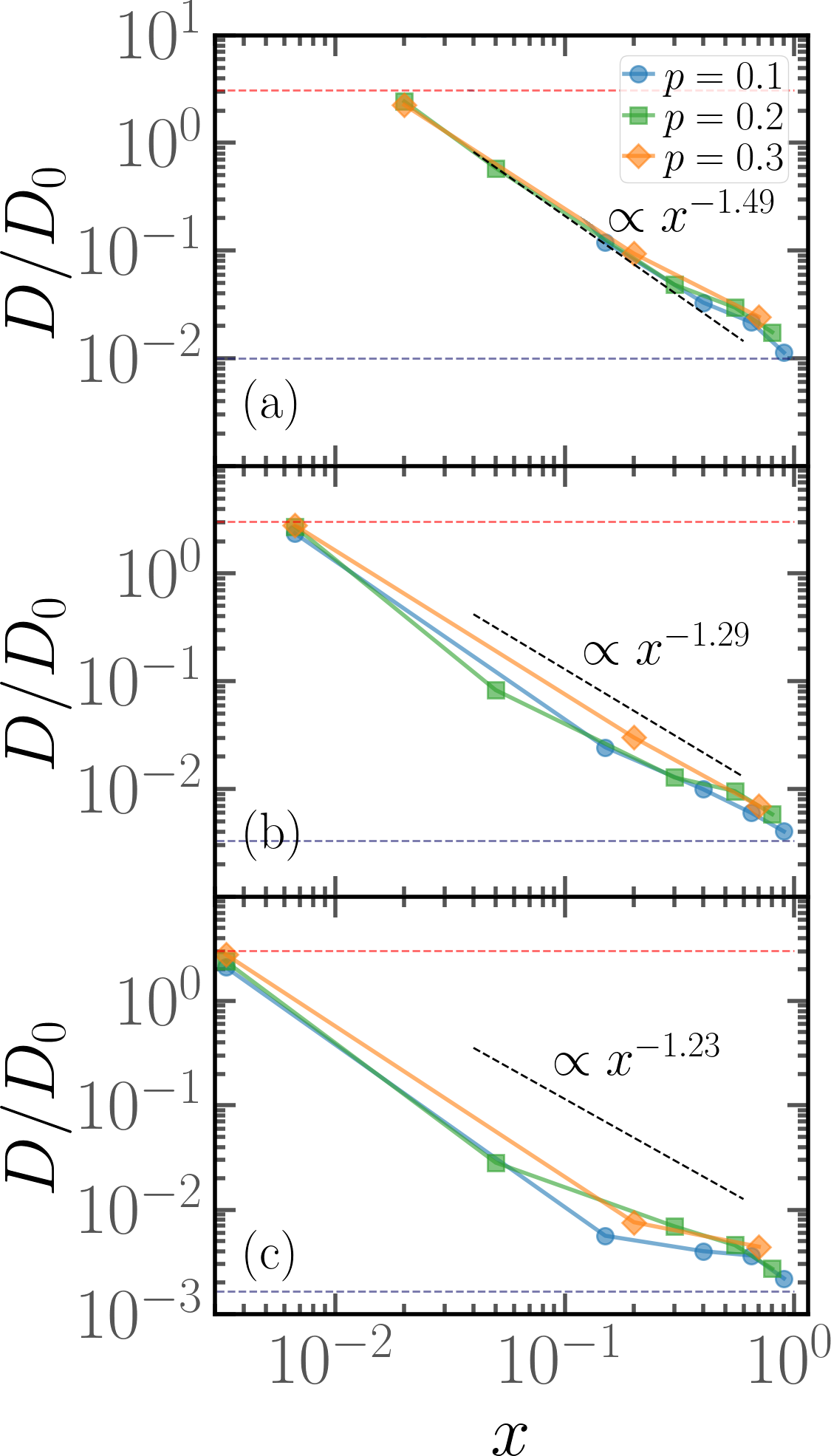}\hfil
	\caption{Diffusion coefficient $D$ of the polymer chains as a function of the parameter $x$ for (a) for $N = 100$; {(b)} for $N = 300$; {(c)} for $N = 600$. The red dashed line refers to the diffusion coefficient of a fully active chain of $\mathrm{Pe}=$10, as predicted by Eq.(6) of Ref.~\cite{bianco2018globulelike}; the blue dashed line refers to the Rouse diffusion coefficient of a passive chain $D_0/N$.}
	\label{fig:diff_Na}
\end{figure}

As hinted by the data reported in Fig.~\ref{fig:msd_long}, the long-time diffusion coefficient $D/D_0$ of partially active linear polymers with one active block depends on the contour position of such active block. We show more in detail this dependency in Fig.~\ref{fig:diff_Na} for polymer chains of length $N=100$ (panel a), $N=$300 (panel b), $N=$600 (panel c), for different values of $p$.

We can observe that, upon increasing $x$, the diffusion coefficient decreases as a power law, whose exponent depends on the size of the polymer. Interestingly, at $x=1/N$ the diffusion coefficient is compatible with the value predicted in Ref.~\cite{bianco2018globulelike} for fully active polymers; conversely, for large values of $x$ the diffusion coefficient becomes compatible with the passive Rouse theory prediction. Thus the contour position modulates the mobility of the chain, at least in the range of values of $p$ considered.   
Further, we can also observe that the data for different values of $p$ roughly fall on the same curve; this happens for polymers of different lengths $N=$100, 300, 600. The data follow a power-law trend; as the curve modulates between the $N$-independent fully active case and the $1/N$ passive case, the resulting power-law exponent depends on $N$. However, this common power law trend indicates that the parameter $x$ captures the effect of the contour position of the active block on the long-term dynamics, regardless of the size of the active block or the size of the chain.\\ 
\begin{figure}[!h]
	\centering
		\includegraphics[width=0.8\columnwidth]{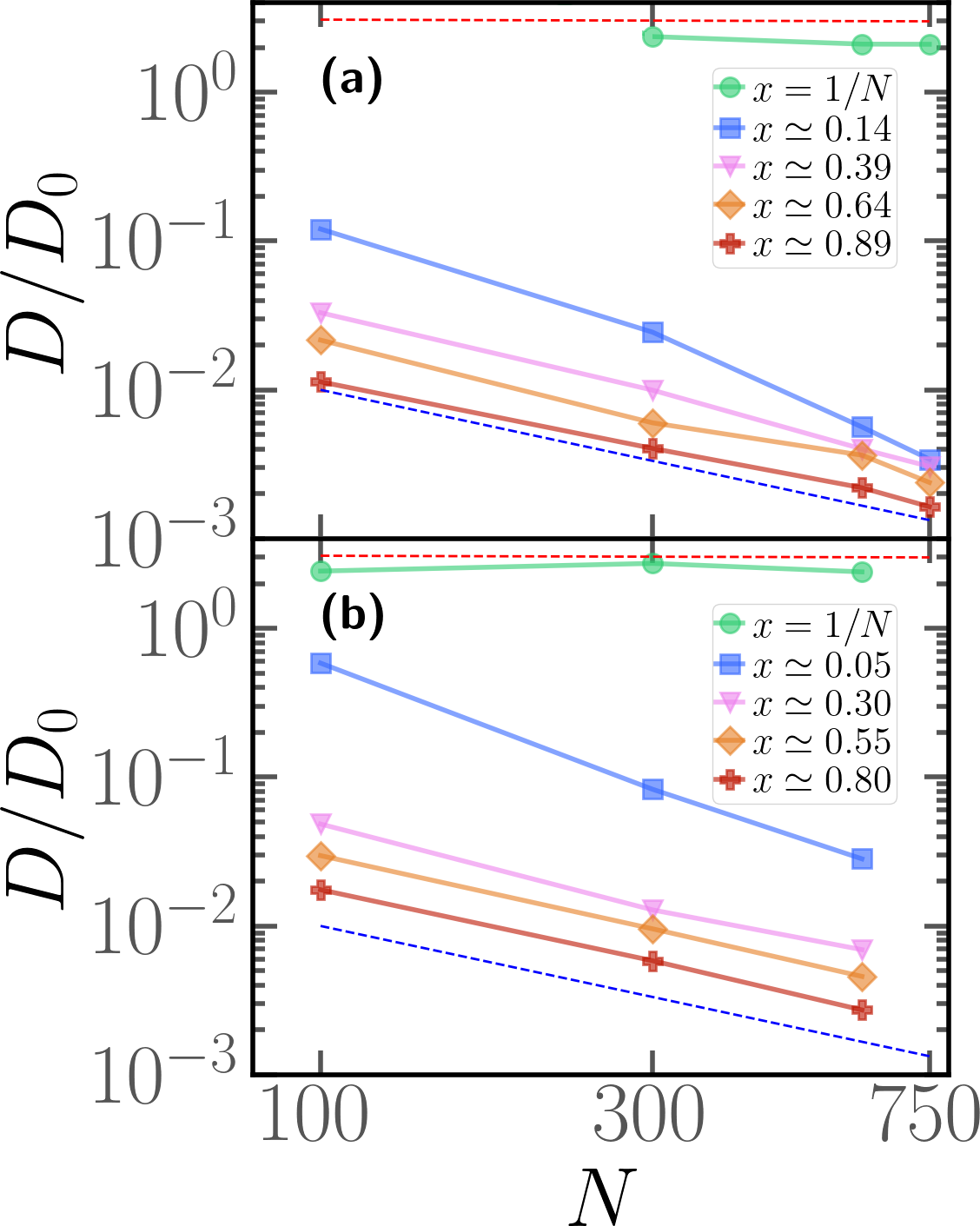}
	\caption{Diffusion coefficient $D/D_0$ as a function of the degree of polymerization $N$ for several values of the parameter $x$ and (a) for $p=0.1$, (b) for $p=0.2$. The red dashed line refers to the diffusion coefficient of a fully active chain of $\mathrm{Pe}=$10, as predicted by Eq.(6) of Ref.~\cite{bianco2018globulelike}; the blue dashed line refers to the Rouse diffusion coefficient of a passive chain $D_0/N$.}
	\label{fig:diff}
\end{figure}

We can recast the same data in a different fashion, by considering the long time diffusion coefficient as a function of the degree of polymerization $N$ for different values of $x$, reported in Fig.~\ref{fig:diff} for $p=$0.1 (panel a), $p=$0.2 (panel b). We see here more clearly that, when $x=1/N$, the diffusion coefficient is compatible with the theoretical prediction of Ref.~\cite{bianco2018globulelike}, valid for a fully active polymer at $\mathrm{Pe}=$10; it is also independent of $N$. The discrepancy between numerical data and prediction is, not surprisingly, smaller at $p=$0.2; however, it is remarkable that the fully active mobility can be achieved, in good approximation, with as low as 1/10 of the original number of active monomers. Further, we observe that the passive-like $1/N$ dependence of the diffusion coefficient is recovered at sufficiently large values of $x$; the value of $x$, at which this happens, seems to decrease upon increasing $p$. However, as also evident from Fig.~\ref{fig:diff_Na}, the diffusion coefficient is a factor of three or more larger than the purely passive case; only for $p=$0.1 and $x=1-p$, i.e. the largest value of $x$ possible, the diffusion coefficient is truly compatible with the passive Rouse prediction. Naturally, upon increasing $p$, the mobility of the chain becomes increasingly larger than the Rouse value even at $x=1-p$, as all the possible arrangements of the $N\cdot p$ active monomers will become increasingly similar, among themselves and with the fully active case.  

\begin{figure}[!h]
	\centering
		\includegraphics[width=0.8\columnwidth]{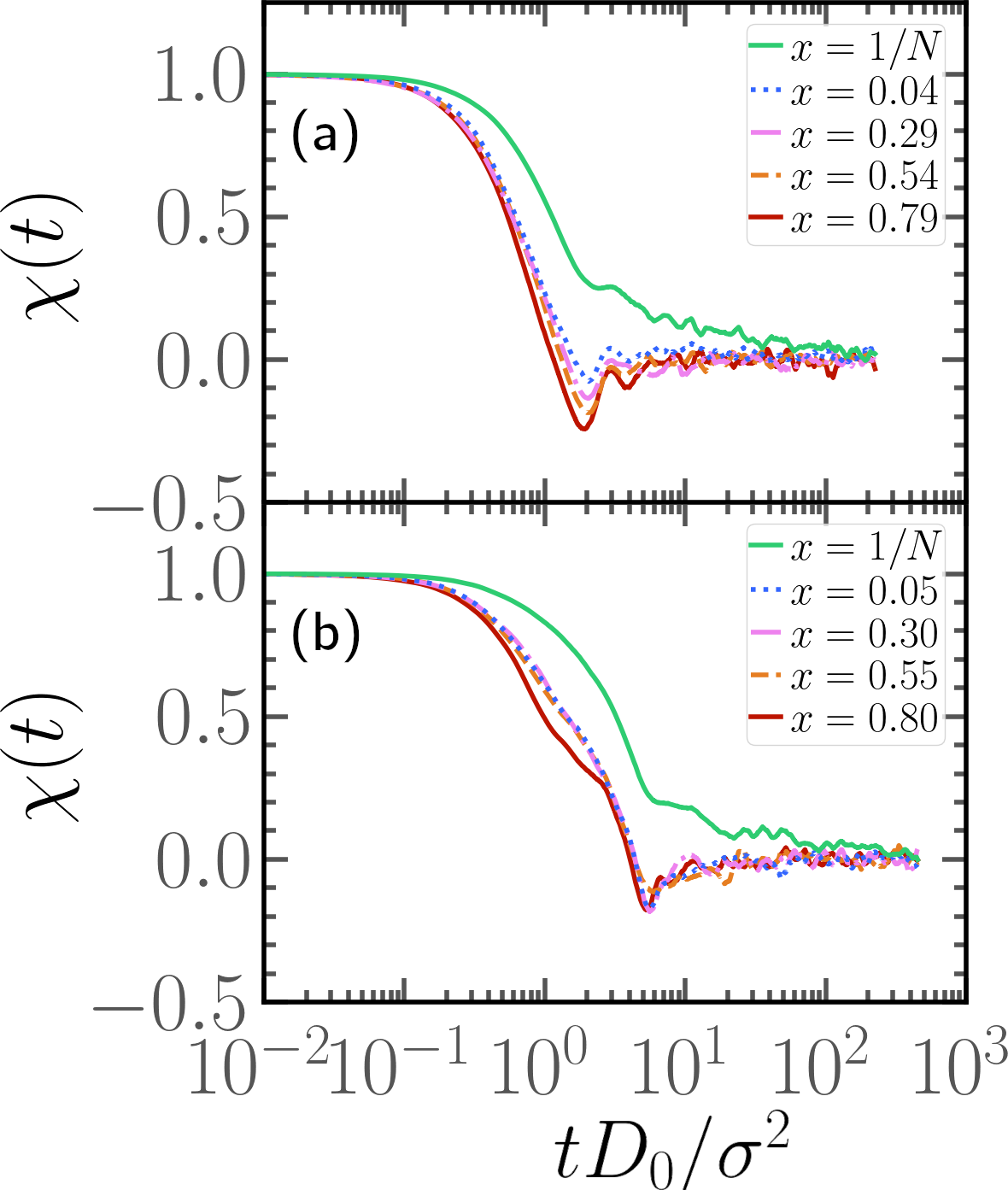}
		\caption{Autocorrelation function of the end-to-end vector of the active block, for different values of $x$ at fixed $\mathrm{Pe} = 10$, $p = 0.2$, and (a) $N = 100$, (b) $N = 300$.}
	\label{fig:autocorrelation}
\end{figure}


In order to understand the dependence of the diffusion coefficient on the position of the active block,  
we consider the autocorrelation function $\chi(t)$ of the end-to-end vector of the active block itself. The function $\chi(t)$ encodes the temporal dependence of the total self-propulsion force that, at such high values of $\mathrm{Pe}$, drives the dynamics; the longer it takes for such total force to decorrelate, the higher will be the diffusion coefficient. We report, in Fig.~\ref{fig:autocorrelation}, the function $\chi(t)$ as a function of the normalized time $t D_0/\sigma^2$, for different values of $x$ at fixed $p=$0.2, $\mathrm{Pe}=$10, and $N=100$ (panel a) and $N=300$ (panel b); data referring to different values of $p$ are reported in the Supplemental Material. We can observe, in both panels, that the autocorrelation function decays more slowly for $x=1/N$ than for higher values of $x$. Thus, the end-to-end vector of the active block and, thus, the total self-propulsion force, maintains the same direction for a longer time when it is as close as possible to the head of the polymer. As soon as a few passive monomers are added in front of the active block, $\chi(t)$ decays more sharply and shows an anti-correlation at some characteristic time. This suggests that the end-to-end vector of the active block tends to point in the opposite direction after a certain characteristic time, in a sort of tumbling motion. Interestingly, the typical time, associated with such ``tumbling'' motion, increases with $N$ and with $p$ (see Supplemental Material).

\subsection{Activity-induced persistence length enhances chain mobility}
\label{sec:persist}
Figures~\ref{fig:diff_Na} and \ref{fig:diff} clearly show that, if a block of active monomers is placed close to the head of the chain, the whole polymer has the same mobility as its fully active counterpart. As mentioned, this result is remarkable and counter-intuitive, since the total active force on the center of mass is as low as 1/10 of its fully active value.

\begin{figure}[!h]
	\centering
	\includegraphics[width=0.8\columnwidth]{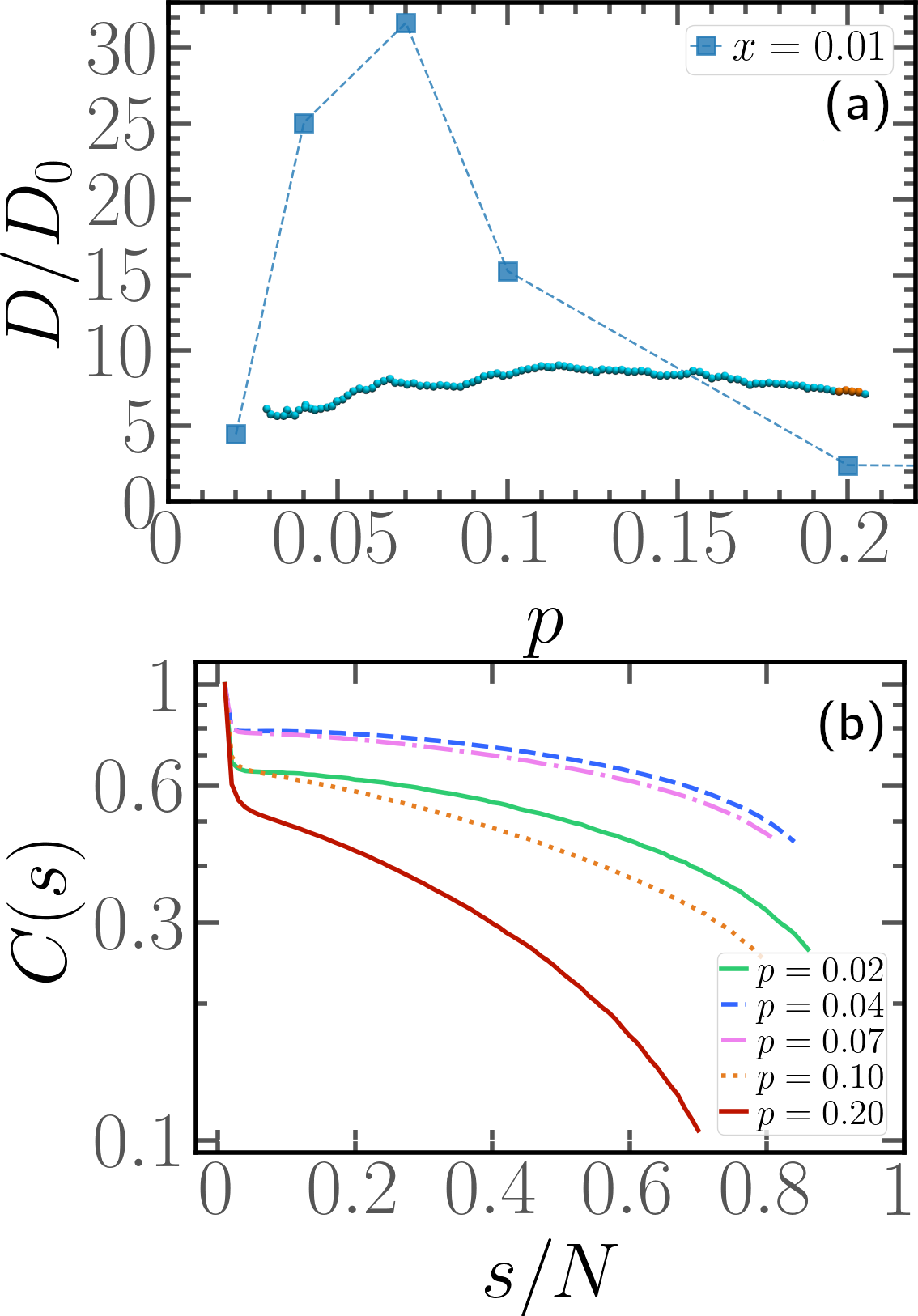}
	\caption{(a) Diffusion coefficient as a function of $p$, for $N = 100$ and $x = 1/N$. Snapshot in the inset refers to a typical conformation observed for $\mathrm{Pe} = 10$, and $p=0.04$. The dotted line is a guide to the eye. (b) Tangent-tangent correlation within the passive section as a function of the contour distance, for $N = 100$, $x = 1/N$, and different values of $p$.}
	\label{fig:last}
\end{figure}

Even more surprising is the fact that, decreasing the percentage of the active monomer further at fixed $x=1/N$, we find a strong enhancement of the long-time diffusion coefficient. This result is reported in Fig.~\ref{fig:last}a, where we show the diffusion coefficient  $D/D_0$ as a function of the fraction of active monomers $p$ for a chain of fixed length $N=$ 100. The results reported, for $p<$0.1, have been obtained from dedicated simulations, averaging over $M=$ 25 independent realizations. We observe that, for $p\leq$0.1, the diffusion coefficient displays a non-monotonic behaviour. Upon increasing $p$, $D/D_0$ first increases, reaching a value up to a factor of 10 larger with respect to the fully active reference case; after the maximum at $p\approx$ 0.07, $D/D_0$ decreases back to the fully active value. Interestingly, the lowest value of $p$ considered, $p=$ 0.02, corresponds to only two active monomers; yet we recover the same diffusion coefficient as for the fully active case. The reason for this non-monotonic behaviour lies in the effective persistence length, observed in active polymers with tangential self-propulsion\cite{bianco2018globulelike}. For very small values of $p$, the size of the active block falls below the persistence length: the active monomers' will thus tend to be aligned and will drag the rest of the chain with them (see snapshot in the inset of Fig.~\ref{fig:last}a), increasing the diffusion coefficient via an increase of the decorrelation time. We remark that, for $N=$100 and $p=$0.1, the active block is short enough to fall into this category; thus the value of $D/D_0$ for $x=1/N$ was not reported neither in Fig.~\ref{fig:diff_Na}a nor in Fig.~\ref{fig:diff}a.
This scenario is confirmed by the end-to-end time correlation $\chi(t)$ for $N=$100, $x=1/N$ and $p=$0.1, reported in the Supplemental Material, which shows an enhancement of the correlation time with respect to the $p=$0.2 case (see Fig.~\ref{fig:autocorrelation}). Further, Fig.~\ref{fig:last}b, shows the tangent-tangent correlation function within the passive section for the same systems as in Fig.~\ref{fig:last}a. We observe that tangent vectors remain correlated over the entire length of the passive chain for the two values of $p$ with the highest mobility. We also observe a quite strong correlation between the tangent-tangent persistence and the diffusion, i.e. when the passive section is most straight, the diffusion is larger. This strengthens the picture of an active polymer with a rod-like conformation, which leads to a longer decorrelation time. Finally, notice that this enhancement will happen, upon increasing $N$, at vanishing $p$; this is a consequence of the fact that the effective persistence length depends on the value of $\mathrm{Pe}$ and not on $N$. 


\section{Conclusions}

In summary, we studied how the arrangement of a certain fraction of active monomers influences the conformation and dynamics of partially active linear polymers. Within the ensemble of all the possible arrangements, we focused on the specific case of a single, contiguous block of active sites. The reason is twofold. First, we showed that, on a population level, random arrangements are all dynamically equivalent; on the contrary, arranging the active sites in contiguous blocks gives rise, in a population of non-interacting chains, to a dynamic heterogeneity. The population sample was constructed by placing the active blocks randomly along the contour; the polymers were, apart from this detail, identical, i.e. same values of $N$, $\mathrm{Pe}$, and $p$. This heterogeneity persists when more than one active block is present; however, we found, in the single block case, that the parameter $x$, given by the minimum contour distance of the first active monomer from the ``head'' of the chain over the total contour length, provides a way to rationalize the conformational and dynamical properties of these partially active chains. In fact, we show that chains characterized by a small value of $x$ are much more elongated and much more mobile with respect to their counterpart characterized by large values of $x$. In other words, the contour position of the active block determines the chain conformation and dynamics; looking at a random population, this explains the emergence of dynamical heterogeneity. Interestingly, conformation and dynamics are here not entirely coupled: while the diffusion coefficient decreases monotonically upon increasing $x$, the shape and size of the chains show a non-monotonicity, the minimum being located when the active block is exactly in the middle of the chain. The contour position also influences the scaling properties: the measured scaling exponent $\nu$ of the gyration radius is $\nu \simeq 1$ for $x=1/N$, while it decreases for larger values of $x$ to a value slightly larger than the passive, self-avoiding reference $\nu=0.588$. In general, the value of $\nu$ will always result from the weighted average between the active block, which tends to be globule-like and the passive sections, which tends to be very extended. Further, in the same perspective, we show that the diffusion coefficient remains $N$-independent, and compatible with the fully active case, only for very small values of $x$; upon adding a few passive monomers between the ``head'' end of the chain and the beginning of the active section, $D$ becomes again a decreasing function of $N$. The increase of the diffusion coefficient at very small values of $x$ can be connected to an anomalous behaviour of the time correlation function of the self-propulsion force, that disappears upon increasing $x$. Finally, this effect can also be exploited and enhanced by \textit{decreasing} the fraction of active sites at $x=1/N$. This very counter-intuitive result is connected with the effective persistence length, induced by the tangential activity; when the active block is very short, the monomers align, causing a collective stretching of the chain that increases its decorrelation time and its diffusion.\\These results show that partially active polymers display a very rich dynamical scenario; understanding their properties will be most useful in guiding the modelisation of filamentous micro-organism and worms. Further, understanding the properties of partially active polymers may also guide the design of artificial, soft robots, optimizing the use of active sections in order to improve control and cost. From a more polymeric perspective, it would be interesting to investigate how the position of a block of active monomers influences the entanglement in very dense conditions\cite{tejedor2023,savoie2023}. Finally, as mentioned in the introduction, chromatin is distinctly characterized by ``sections'', among which the active one is bound to be out-of-equilibrium due to the action of ATP-driven molecular motors: understanding the phenomenology of polymer chains with active sections will be important, in order to include the non-equilibrium effects in chromatin models.

\section*{Author Contributions}
Conceptualization, E.L., S.K. and M.V.; investigation, E.L. and M.V.; formal analysis, M.V.; writing – original
draft preparation, M.V.; writing – review and editing, E.L. and S.K.; supervision, E.L.; funding acquisition, E.L.
All authors have read and agreed to the published version of the manuscript.

\section*{Conflicts of interest}
There are no conflicts to declare.

\section*{Acknowledgements}
This work has been funded by the project ``LOCA\_BIRD2222\_01'' of the University of Padova. E. L. acknowledges support from the MIUR grant Rita Levi Montalcini. The computational results presented have been achieved using the Vienna Scientific Cluster (VSC); CloudVeneto is also acknowledged for the use of computing and storage facilities. S.K. acknowledges financial support from PNRR Grant CN\_00000013\_CN-HPC, M4C2I1.4, spoke 7, funded by NextGenerationEU.


\providecommand*{\mcitethebibliography}{\thebibliography}
\csname @ifundefined\endcsname{endmcitethebibliography}
{\let\endmcitethebibliography\endthebibliography}{}

\pagebreak

\pagestyle{fancy}
\thispagestyle{plain}
\fancypagestyle{plain}{
\renewcommand{\headrulewidth}{0pt}
}

\makeFNbottom
\makeatletter
\renewcommand\LARGE{\@setfontsize\LARGE{15pt}{17}}
\renewcommand\Large{\@setfontsize\Large{12pt}{14}}
\renewcommand\large{\@setfontsize\large{10pt}{12}}
\renewcommand\footnotesize{\@setfontsize\footnotesize{7pt}{10}}
\makeatother

\renewcommand{\thefootnote}{\fnsymbol{footnote}}
\renewcommand\footnoterule{\vspace*{1pt}%
\color{cream}\hrule width 3.5in height 0.4pt \color{black}\vspace*{5pt}} 
\setcounter{secnumdepth}{5}

\makeatletter 
\renewcommand\@biblabel[1]{#1}            
\renewcommand\@makefntext[1]%
{\noindent\makebox[0pt][r]{\@thefnmark\,}#1}
\makeatother 
\renewcommand{\figurename}{\small{Fig.}~}
\sectionfont{\sffamily\Large}
\subsectionfont{\normalsize}
\subsubsectionfont{\bf}
\setstretch{1.125} 
\setlength{\skip\footins}{0.8cm}
\setlength{\footnotesep}{0.25cm}
\setlength{\jot}{10pt}
\titlespacing*{\section}{0pt}{4pt}{4pt}
\titlespacing*{\subsection}{0pt}{15pt}{1pt}

\fancyfoot{}
\fancyfoot[LO,RE]{\vspace{-7.1pt}\includegraphics[height=9pt]{LF}}
\fancyfoot[CO]{\vspace{-7.1pt}\hspace{13.2cm}\includegraphics{RF}}
\fancyfoot[CE]{\vspace{-7.2pt}\hspace{-14.2cm}\includegraphics{RF}}
\fancyfoot[RO]{\footnotesize{\sffamily{1--\pageref{LastPage} ~\textbar  \hspace{2pt}\thepage}}}
\fancyfoot[LE]{\footnotesize{\sffamily{\thepage~\textbar\hspace{3.45cm} 1--\pageref{LastPage}}}}
\fancyhead{}
\renewcommand{\headrulewidth}{0pt} 
\renewcommand{\footrulewidth}{0pt}
\setlength{\arrayrulewidth}{1pt}
\setlength{\columnsep}{6.5mm}
\setlength\bibsep{1pt}

\makeatletter 

\makeatother

\twocolumn[
  \begin{@twocolumnfalse}
{\includegraphics[height=30pt]{journal_name}\hfill\raisebox{0pt}[0pt][0pt]{\includegraphics[height=55pt]{RSC_LOGO_CMYK}}\\[1ex]
\includegraphics[width=18.5cm]{header_bar}}\par
\vspace{1em}
\sffamily
\begin{tabular}{m{4.5cm} p{13.5cm} }
%
%
%
%
\includegraphics{DOI} & \noindent\LARGE{\textbf{Conformation and dynamics of partially active linear polymers: Supplemental Information}} \\
\vspace{0.3cm} & \vspace{0.3cm} \\

& \noindent\large{Marin Vatin,$^{\ast}$\textit{$^{a,b}$} Sumanta Kundu,\textit{$^{a,b,c}$} and Emanuele Locatelli\textit{$^{a,b}$}} \\

\end{tabular}

\end{@twocolumnfalse} \vspace{0.6cm}
]
  

\renewcommand*\rmdefault{bch}\normalfont\upshape
\rmfamily
\section*{}
\vspace{-1cm}


\footnotetext{ Dipartimento di Fisica e Astronomia, Universit{\`a} di Padova, via Marzolo 8, I-35131 Padova, Italy \textit{$^{b}$ }INFN, Sezione di Padova, via Marzolo 8, I-35131 Padova, Italy}




\section{Number of arrangements of the active monomers: theoretical calculation}
\label{sec:theocount}

\begin{figure}[!h]
	\centering
	\includegraphics[width=0.8\columnwidth]{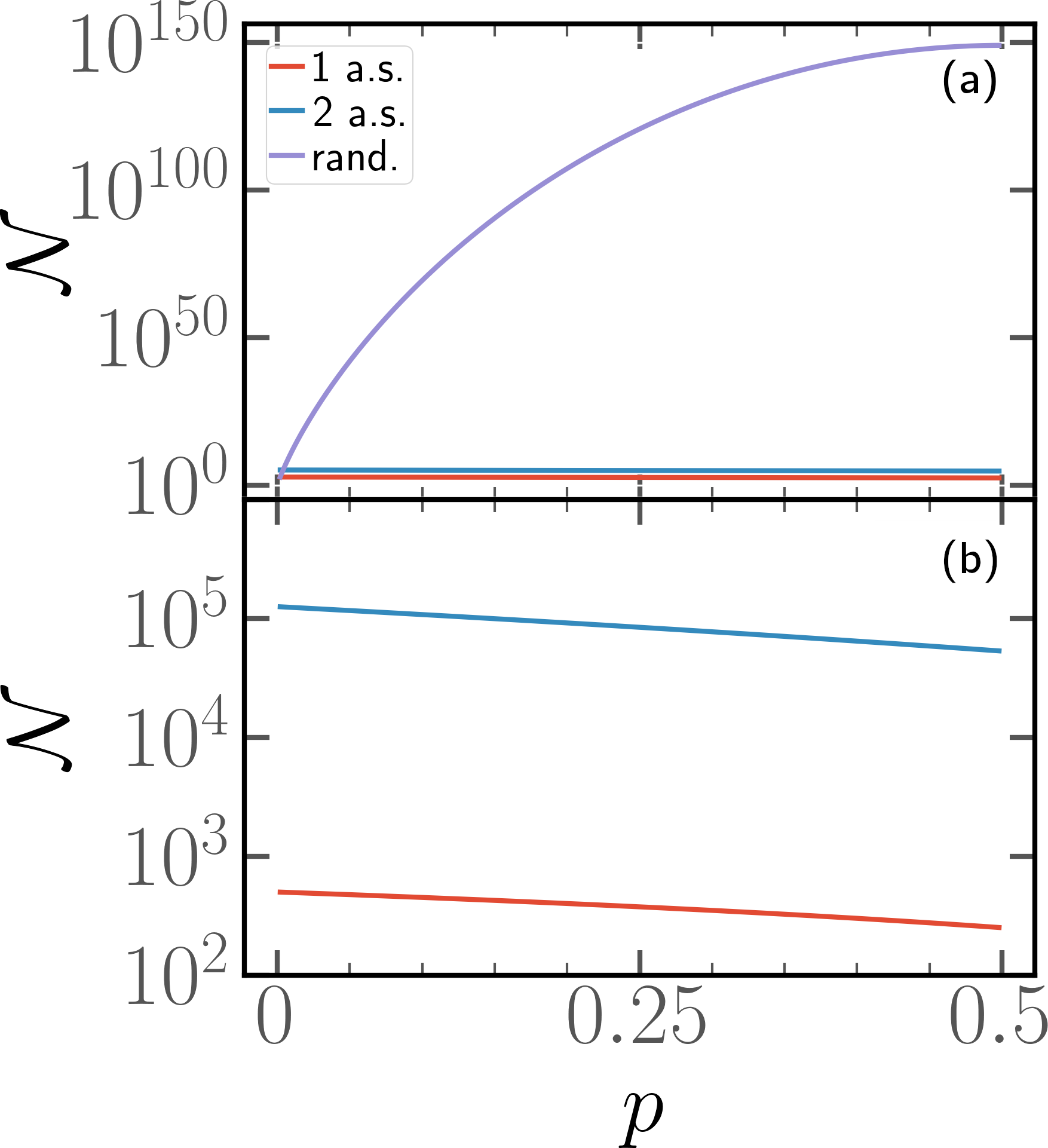}
	\caption{Number of arrangements $\mathcal{N}$ of $N \cdot p$ active monomers (with $N=$500) along the polymer chain as a function of the percentage of active monomers $p$.}
	\label{fig:nb_config}
\end{figure}

We report here on the theoretical estimate of the number of possible arrangements of the active monomers as a function of the their fraction $p$. We recall that we considered four different settings in the main text: a) the active monomers are distributed at random along the chain, b) the active monomers are arranged in one active block, c) the active monomers are arranged in two, non-overlapping active blocks, d) the active monomers are arranged in three, non-overlapping active blocks. We will provide here estimates for settings a)-c).\\
The number of possible arrangements of $N \cdot p$ active monomers within a total of $N$ monomers can be calculated using combinatorial methods. In the random case, the number of configurations $\mathcal{N}$ is simply given by the binomial coefficient:

\begin{equation}
\mathcal{N} = \frac{N!}{(N \cdot (1-p))! \cdot (N \cdot p)!}
\end{equation}

where both $N \cdot p$ and $N \cdot (1-p)$ are integers or should be rounded to the closest integer. 
The set of arrangements where the active monomers are organized in a single block of $N \cdot p$ monomers is a subset of the random case; the number of configurations is simply given by:

\begin{equation}
\mathcal{N}_{1}(N,p) = N \cdot (1-p-1/N) \underset{N \gg 1}{\sim} N \cdot (1-p)  
\end{equation}

where $N \cdot (1-p)$ is the number of inactive monomers; the additional $1/N$ accounts for fact that the first monomer is always passive. The approximation is valid in the limit of very large chains.\\Finally, in the case of two identical blocks, totalling $N \cdot p$ active monomers, the number of arrangements can be approximated as

\begin{equation}
	\mathcal{N}_{2} = \frac{\mathcal{N}_{1}\left(N \cdot p/2,p/2\right) \cdot \mathcal{N}_{1}\left(N \cdot (1-p)/2,p/2\right)}{2}
\end{equation}

where the factor $1/2$ accounts for the fact that the order of the blocks does not matter. Note, however, that the formula is based on the approximation that the blocks are independent, i.e. all positions are possible and may overlap. This approximation remains reasonable as long as the percentage of active monomers remains low and, in any case, overestimates the total number of arrangements. The theoretical predictions are shown in Fig.~\ref{fig:nb_config}.

As shown, the number of arrangements is overwhelmingly larger in the random case than in the one block or two blocks cases. Thus, albeit blocks are possible random arrangements, they are so rare that their effect is negligible at the population level.

\section{Gyration radius, asphericity and prolateness}

In this section, we report complementary data on the size and shape of polymer chains: we report data for polymers of different contour length and P\'eclet number with respect to the main text. 

\begin{figure}[!h]
	\centering
		\includegraphics[width=0.48\columnwidth]{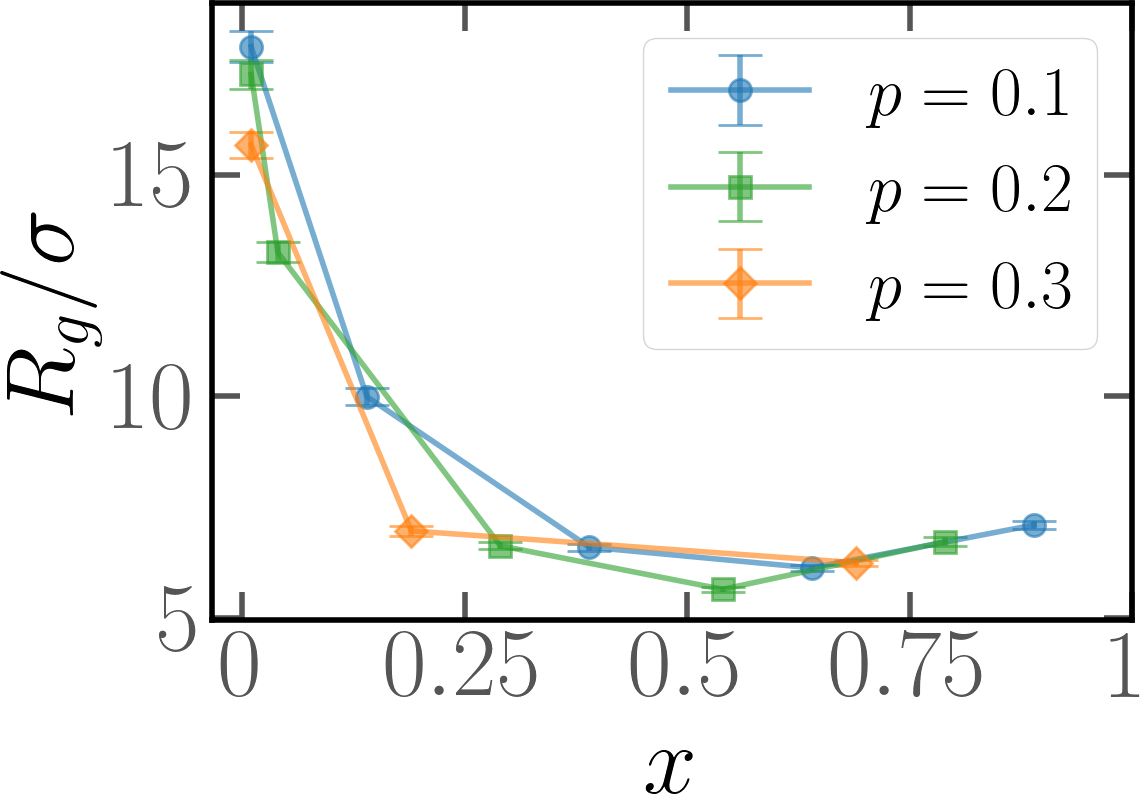}
		\includegraphics[width=0.48\columnwidth]{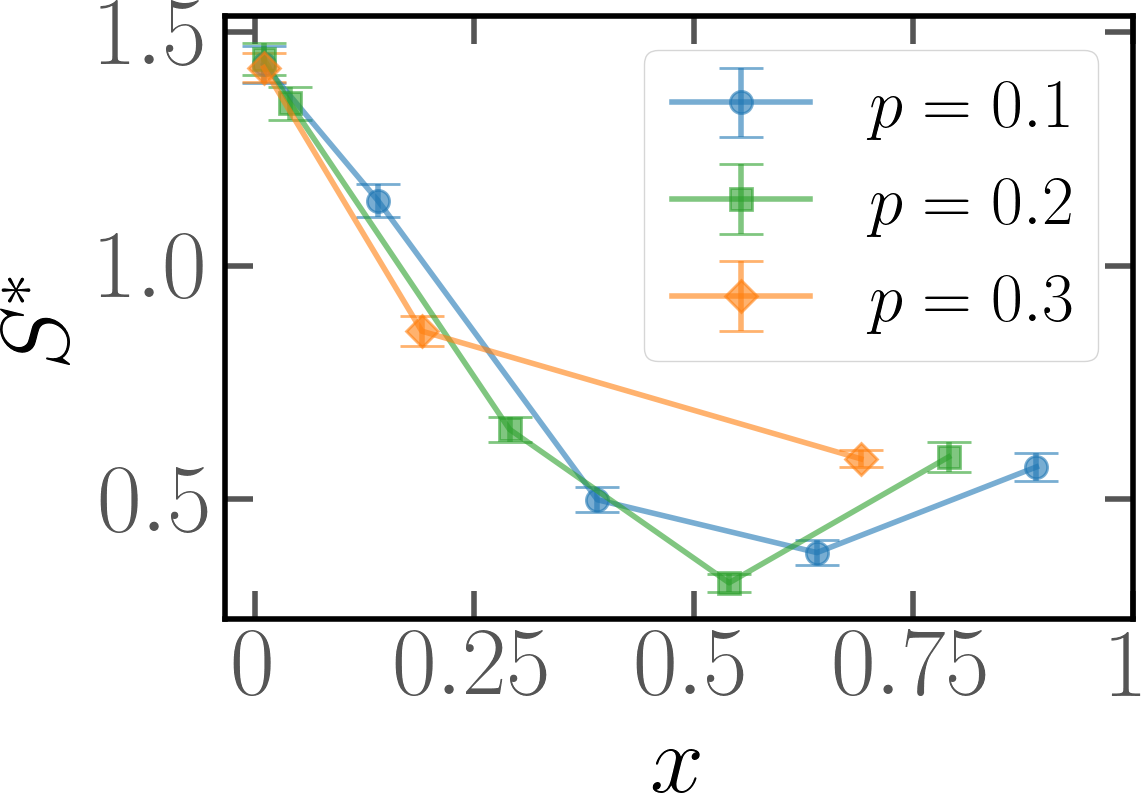}\\
        \includegraphics[width=0.48\columnwidth]{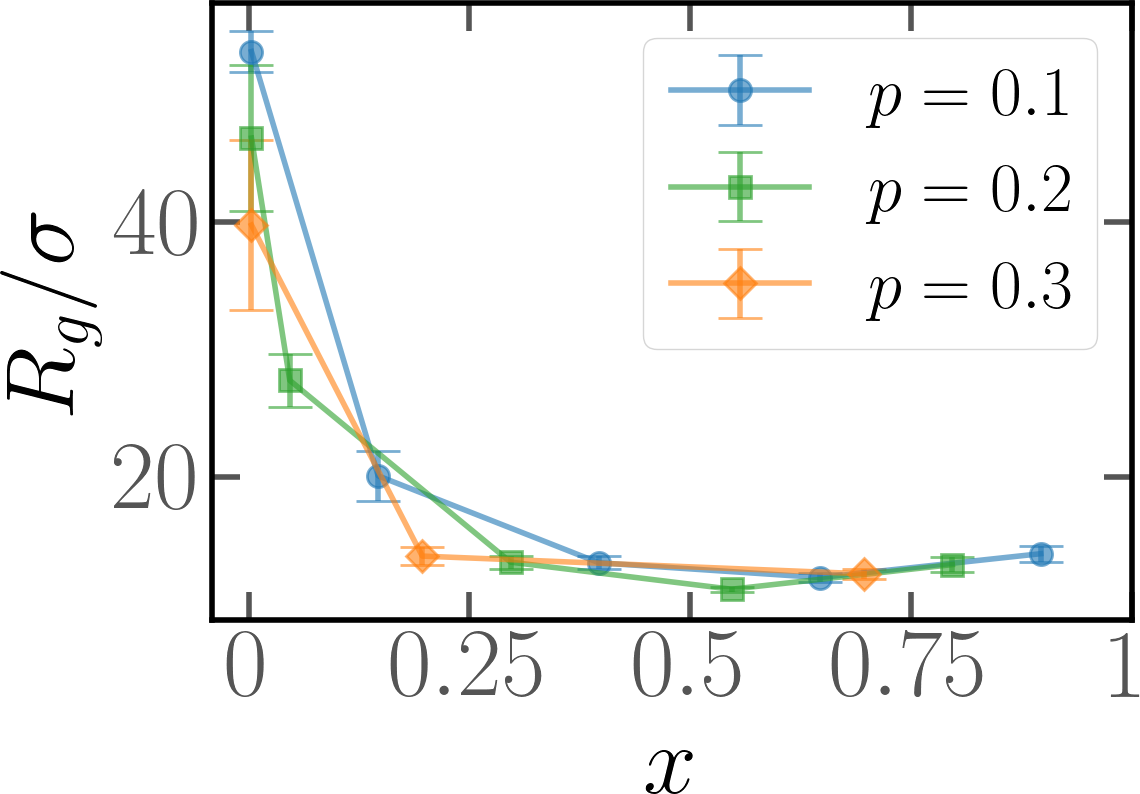}
        \includegraphics[width=0.48\columnwidth]{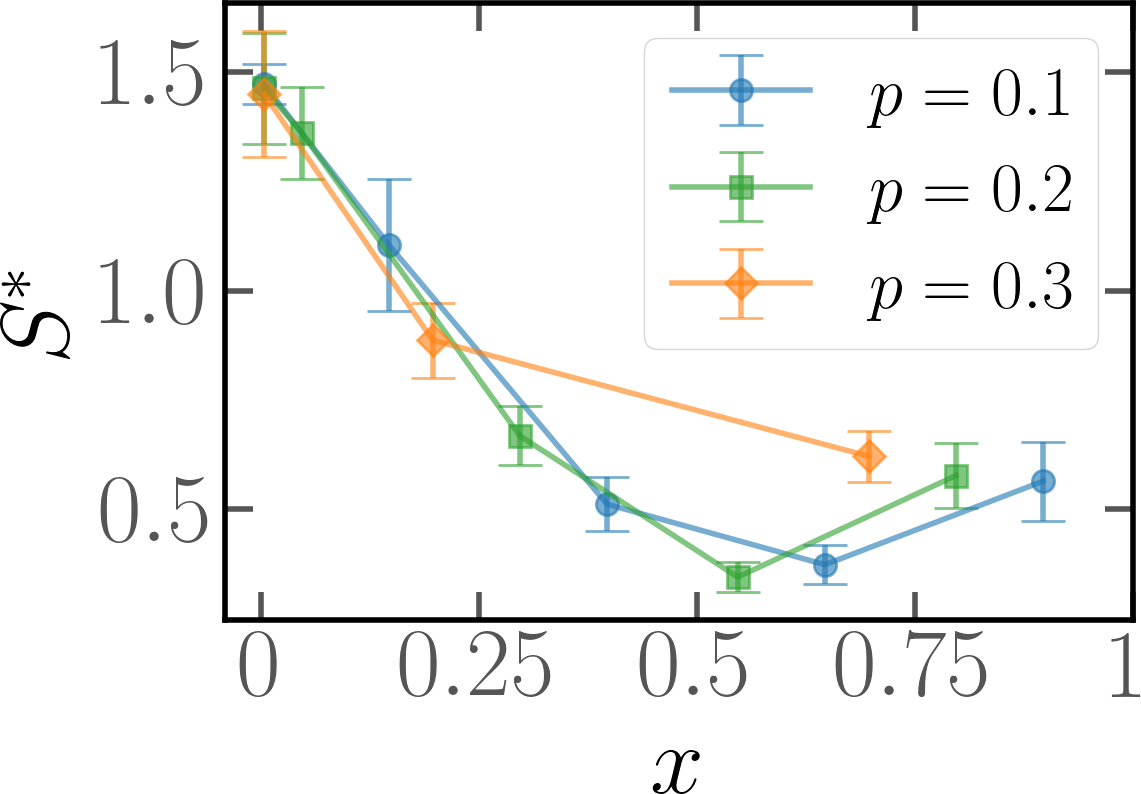}
        \includegraphics[width=0.48\columnwidth]{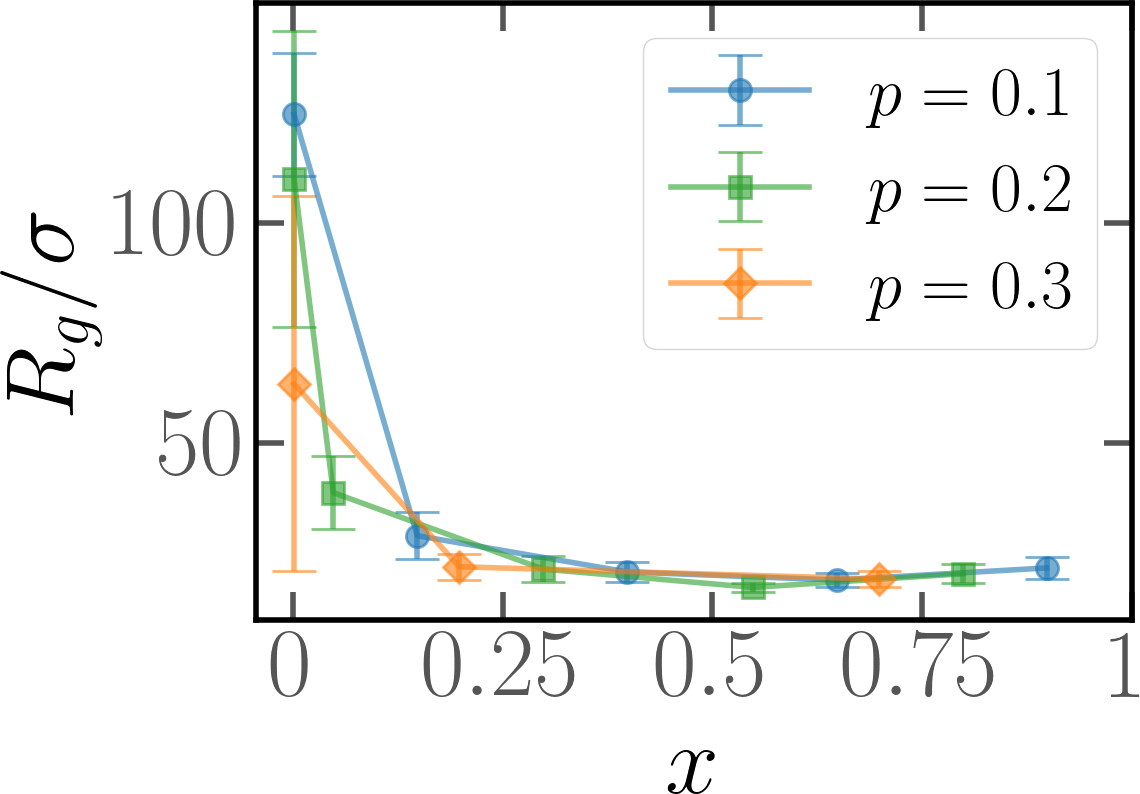}
        \includegraphics[width=0.48\columnwidth]{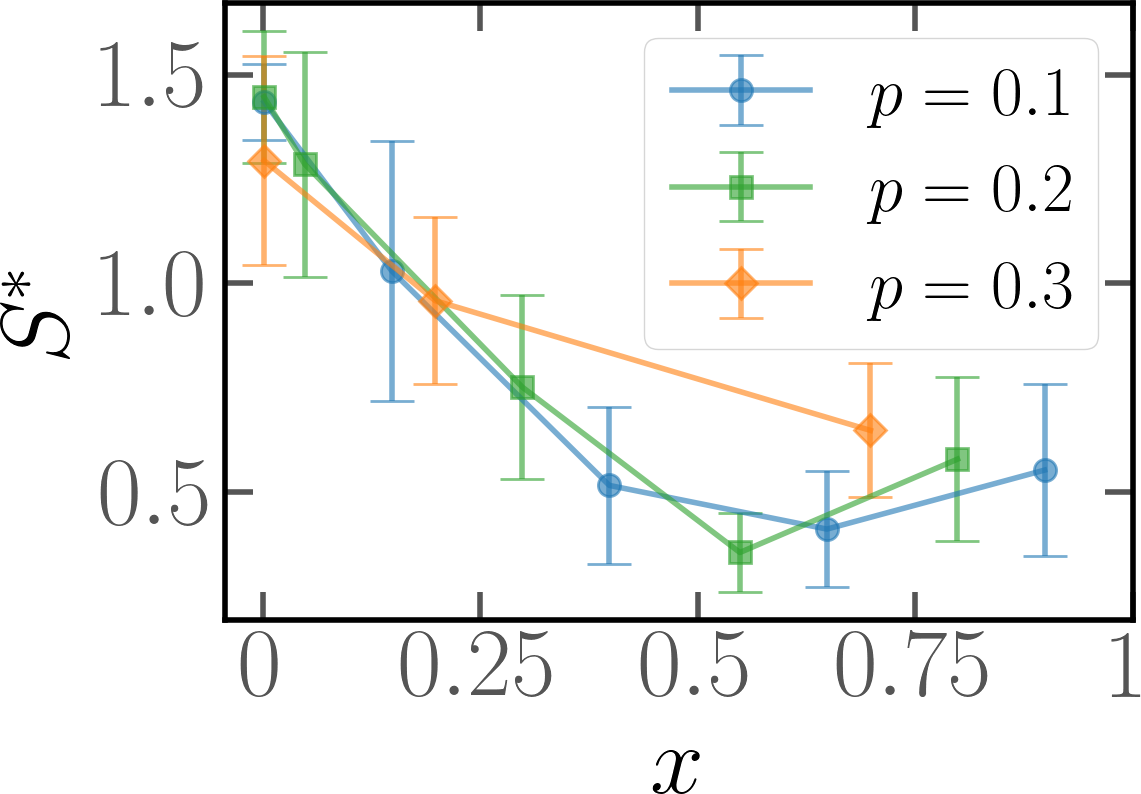}
	\caption{Mean Gyration Radius (left), and prolateness (right) as a function of the parameter $x$ for $\mathrm{Pe}=$10, different values of $p$ and $N = 100$, $N = 300$ $N = 600$. }
	\label{fig:geo100}
\end{figure}

In Fig.~\ref{fig:geo100}, we report the mean gyration radius and prolateness as a function of the parameter $x$ for chains of different degree of polymerization $N=$100,300,600. We can observe that the features, reported in the main text for $N=$500, are visible here as well. Here, no smoothing has been applied to the data; error bars refer to the standard error of the mean over $M=$25 independent identical realizations. This confirms that the general features reported in the main text remain valid for polymers of different length, bearing in mind that, for the values of $p$ and $x$ highlighted in the main text, the scaling breaks due to self-entanglements. 

		
		

		
		

\begin{figure}[!h]
	\centering
		\includegraphics[width=5.8cm]{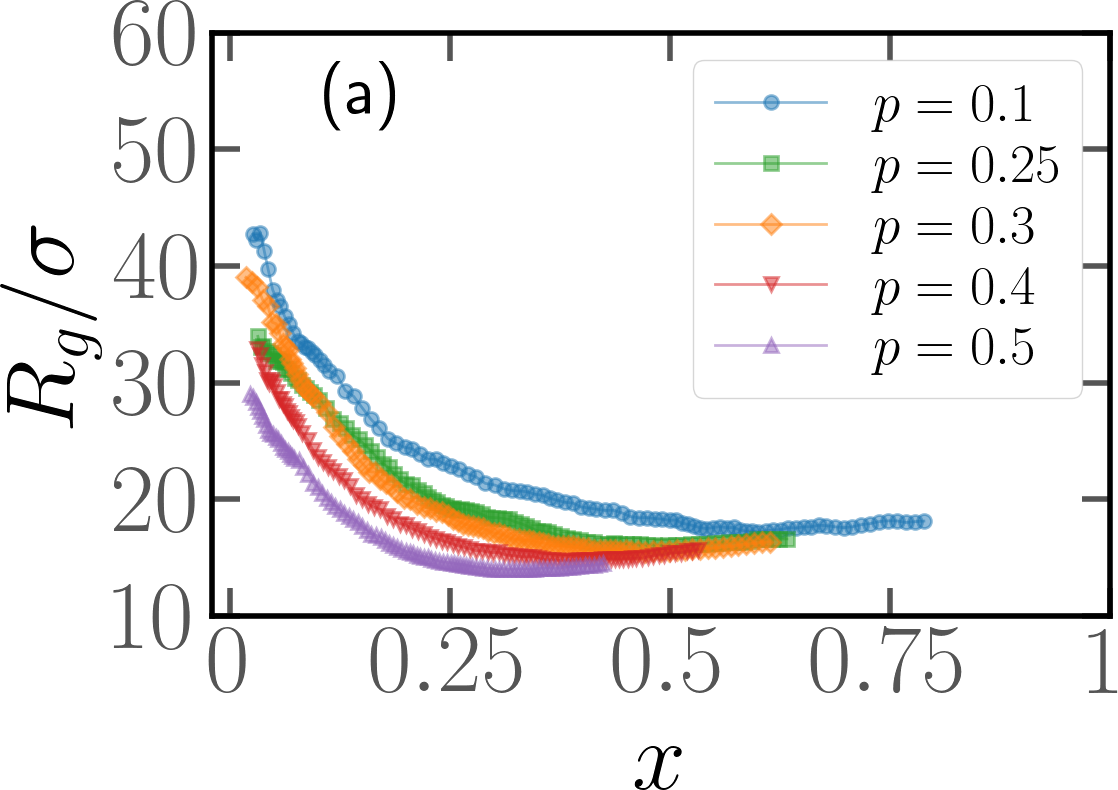}
		\includegraphics[width=5.8cm]{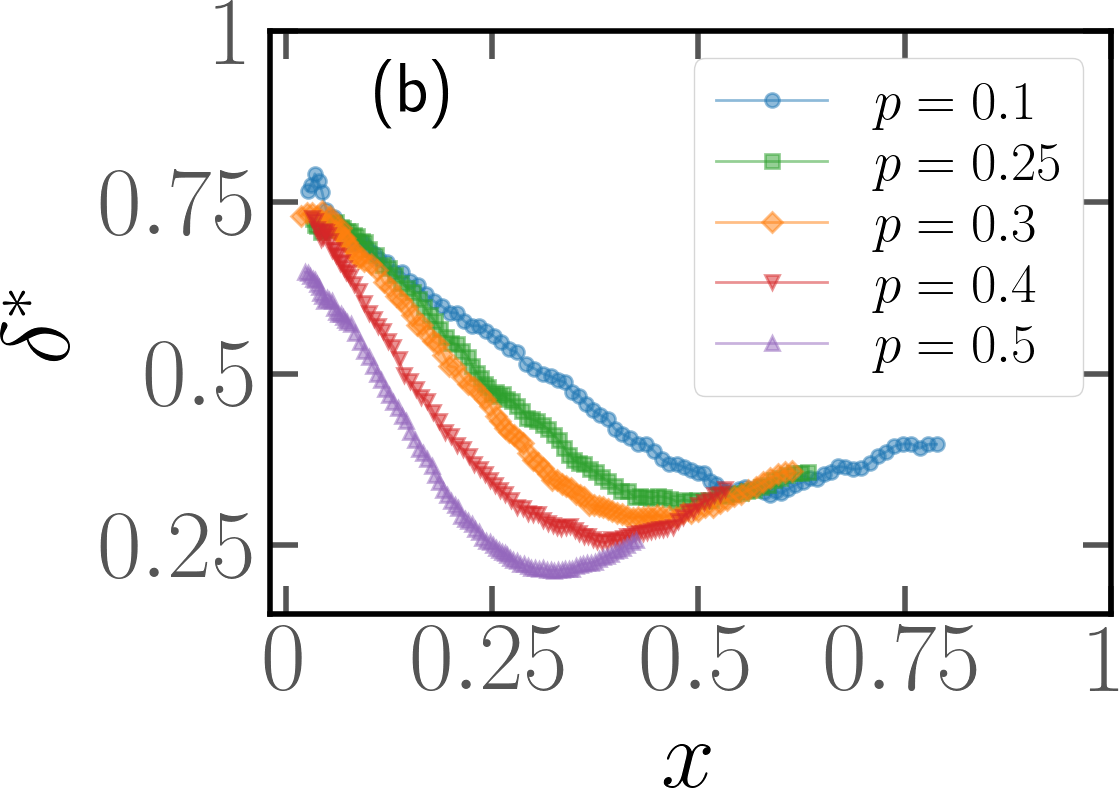}
		\includegraphics[width=5.8cm]{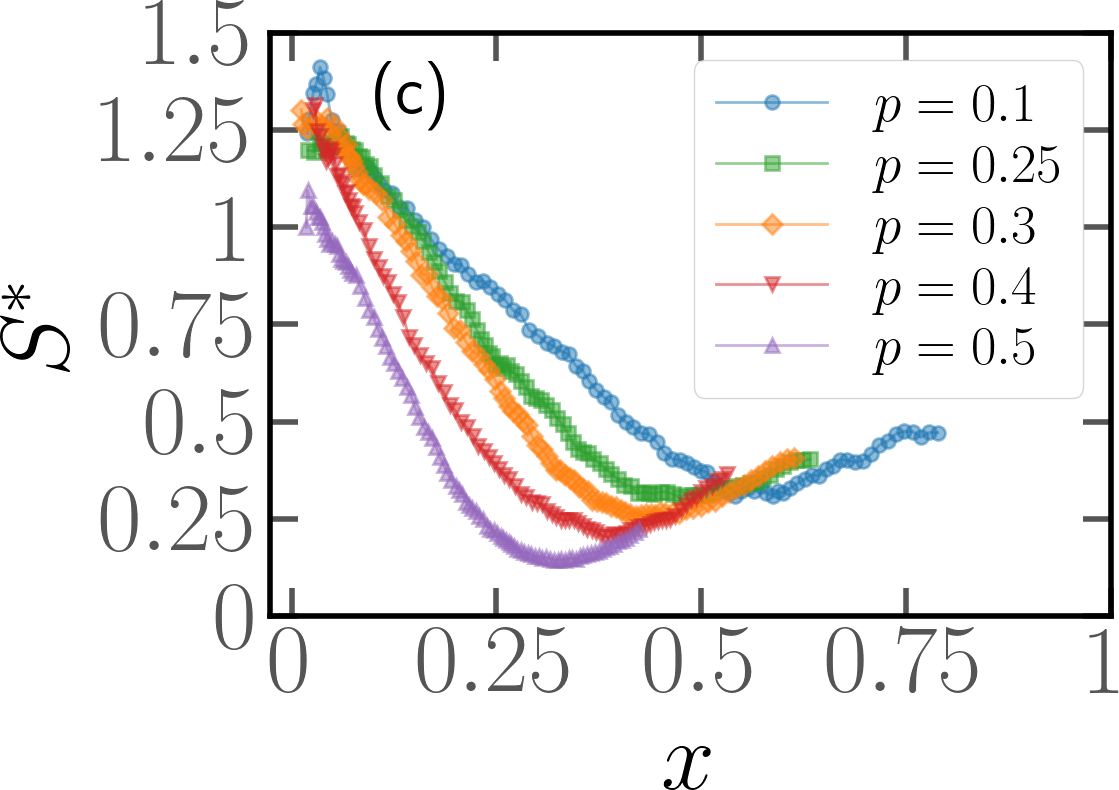}
	 \caption{(a) Gyration Radius, (b) asphericity and (c) prolateness as a function of $x$, at fixed $\mathrm{Pe}=$0.1 for different values of $p$. }
	\label{fig:geo_q_1}
\end{figure}

In Fig.~\ref{fig:geo_q_1} we show the mean gyration radius, asphericity and prolateness as a function of the parameter $x$ for chains of length $N=$500 at fixed $\mathrm{Pe}=$0.1 and different values of $p$. We thus check whether the contour position of the active section influences the conformation and the shape of the chain even at low activity. Indeed, we find that the values of the shape parameters are in semi-qualitative agreement with the values reported in the main text at $\mathrm{Pe}=$10. This means that, even if the individual monomers have a low activity, the effects of the activity on the conformation of the chain can be relevant. Indeed, the expected value of $R_g$, at $Pe=$0, is $R_g/\sigma \approx 0.59 \cdot N^{0.588} = $ 20.21 (for the Kremer-Grest model, we take the scaling prediction from the Supplemental Material of ref.~\cite{locatelli2023nonmonotonous}); thus $R_g$ increases almost by a factor of two at $x=1/N$ and shrinks by roughly 30\% at $x=(1-p)/2$. On the contrary, for a fully active polymer, the relative difference is, at $\mathrm{Pe}=$0.1, of the order of a few percent. However, as suggested by the small values of the non-Gaussian parameter reported in the main text, the effect of the contour position of the active block on the polymer dynamics at small values of $\mathrm{Pe}$ is not equally relevant.

\section{End-to-end autocorrelation function}

In this section, we report complementary data on the time correlation function of the end-to-end vector of the active section at fixed $\mathrm{Pe}=$10. 

\begin{figure}[ht]
	\centering
		\includegraphics[width=0.8\columnwidth]{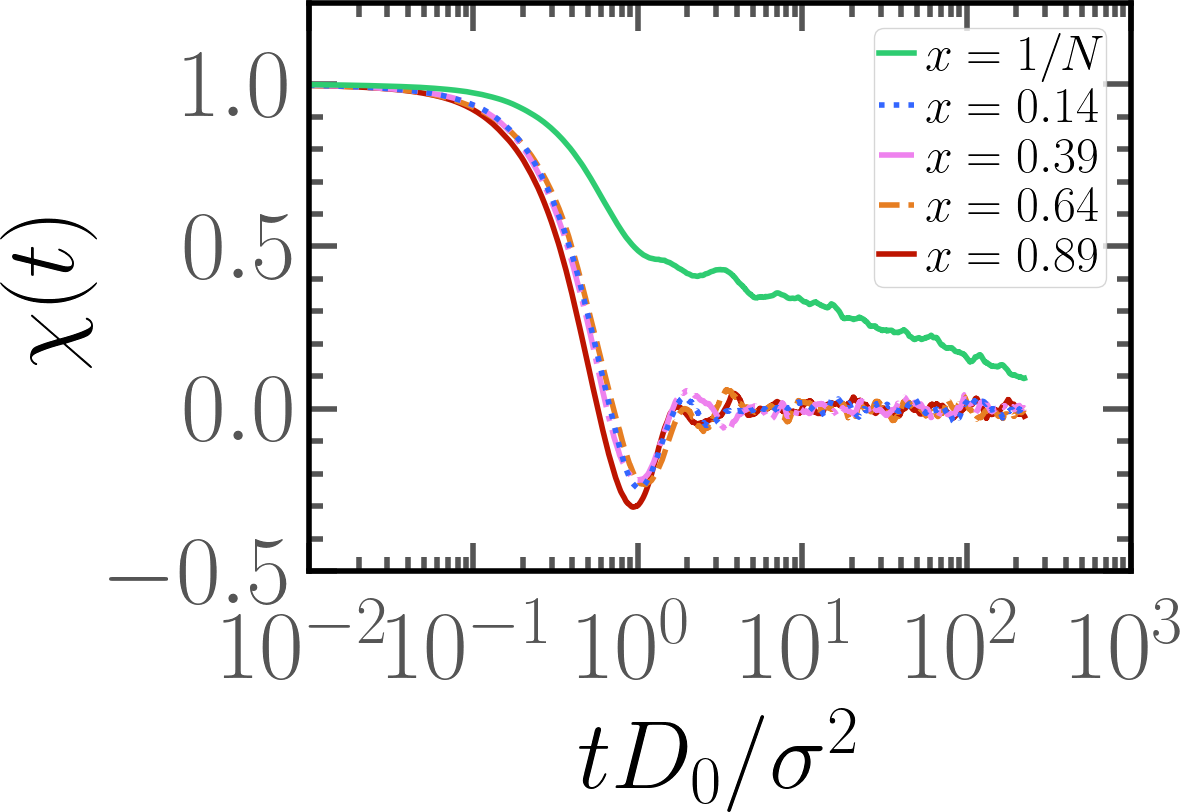}
		\includegraphics[width=0.8\columnwidth]{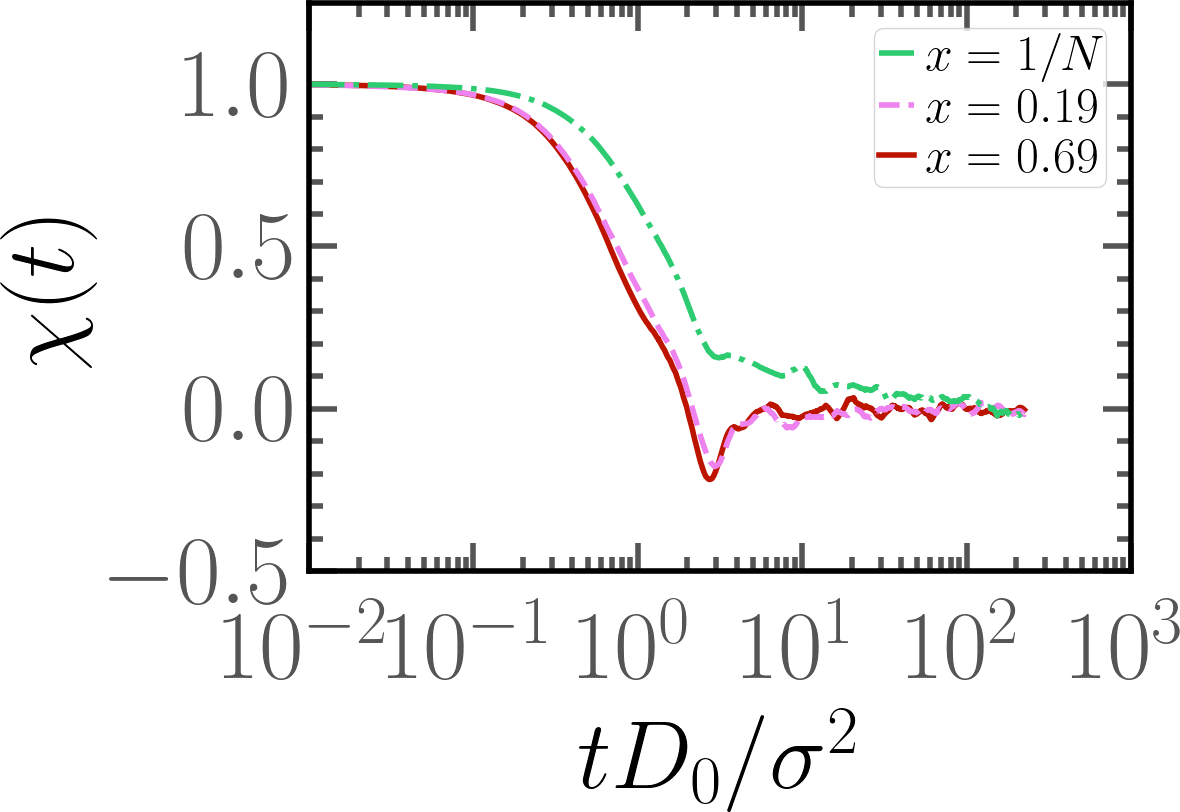}\hfil
	\caption{Time correlation function of the end-to-end vector of the active section at fixed $\mathrm{Pe}=$10, $N=100$ for different values of $x$ and (a) $p =$ 0.1 (b), $p=$0.3. }
\label{fig:autocorr100}
\end{figure}

In Fig.~\ref{fig:autocorr100} we report the time correlation function of the end-to-end vector of the active section as a function of the rescaled time $t D_0/\sigma^2$ at fixed $N=$100 and $\mathrm{Pe}=$10 for different values of $x$. We show in panel {(a)} data referring to $p = $ 0.1 and in panel {(b)} data referring to $p =$ 0.3. We highlight that the anomalous behaviour of the time correlation function is much more evident in panel (a) than in panel (b) or in the data shown in the main text. For $N=$100, $p=$0.1 the active section takes much longer time to decorrelate, to the point that its value only approaches zero at the end of the available simulation window. The tail of the correlation function looks, in the semi-log representation of the plot, almost linear; this would suggest a very slow logarithmic decay. As mentioned in the main text, this is a consequence of the effective persistence length, connected to the tangential activity.

\begin{figure}[!t]
	\centering
		\includegraphics[width=0.8\columnwidth]{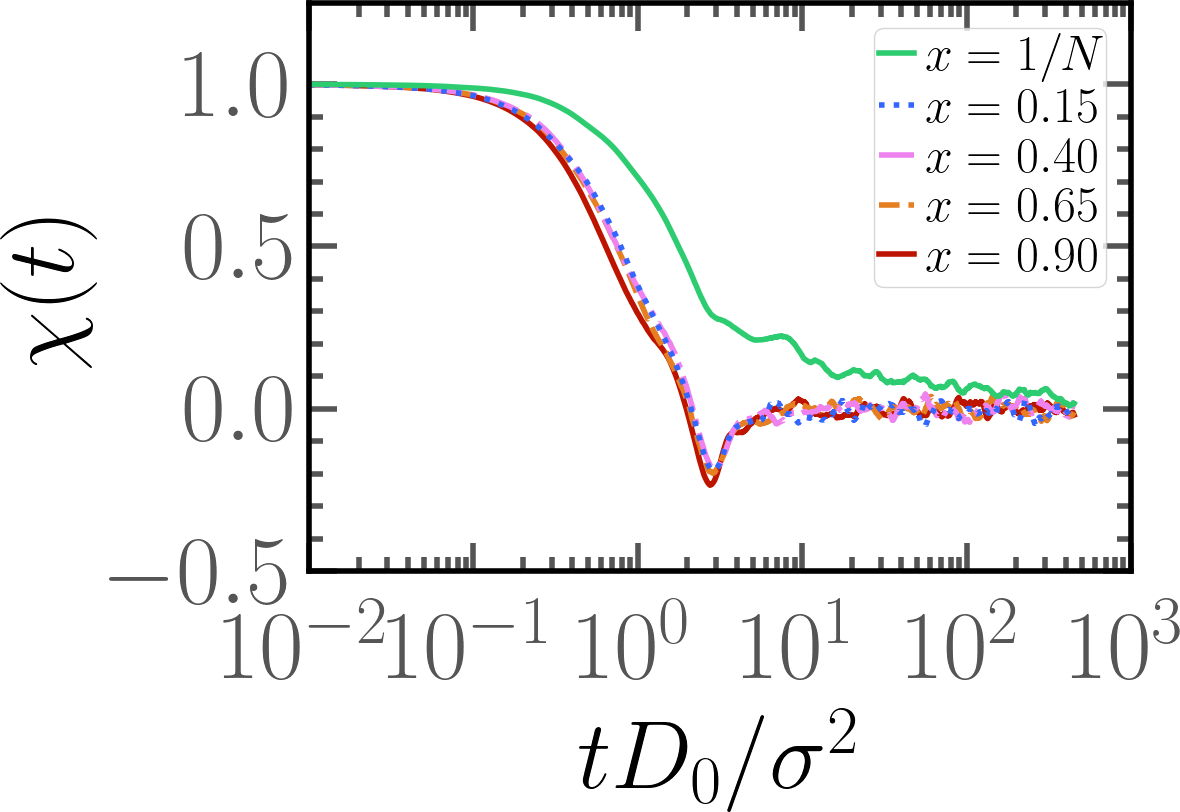}
		\includegraphics[width=0.8\columnwidth]{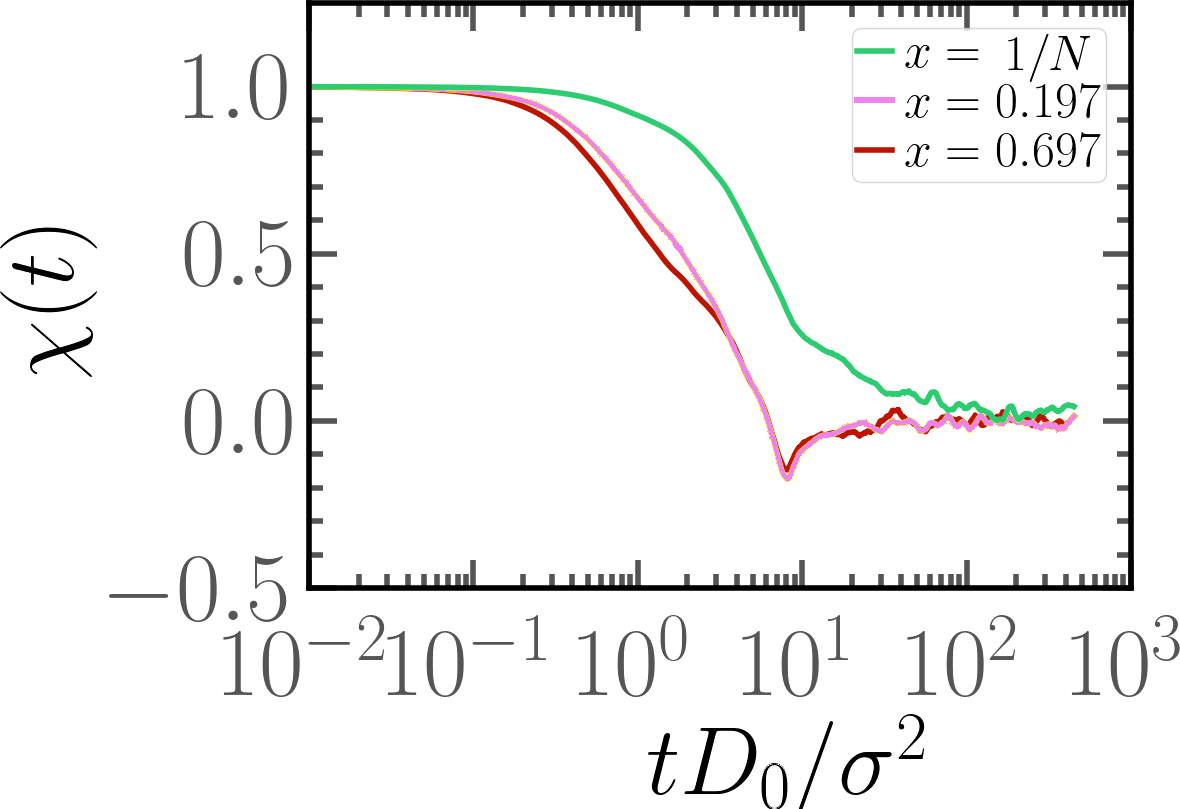}
	\caption{Time correlation function of the end-to-end vector of the active section at fixed $\mathrm{Pe}=$10, $N=100$ for different values of $x$ and (a) $p =$ 0.1 (b), $p=$0.3. }
 \label{fig:autocorr300}
\end{figure}

In Fig.~\ref{fig:autocorr300} we report the time correlation function of the end-to-end vector of the active section as a function of the rescaled time $t D_0/\sigma^2$ at fixed $N=$300 and $\mathrm{Pe}=$10 for different values of $x$. We show in panel {(a)} data referring to $p = $ 0.1 and in panel {(b)} data referring to $p =$ 0.3. Comparing to the previous figure, we notice that that the anomalous behaviour of the time correlation function at $x=1/N$ is less extreme for $N=$300, again simply because the length of the active block increases upon increasing $N$ at the same value of $p$. As in the main text, we notice in both Figs.~\ref{fig:autocorr100}, ~\ref{fig:autocorr300} the characteristic time, associated with ``tumbling'' motion, increases upon increasing both $N$ and $p$.

\section{Multiple active blocks: configurational properties}

In this section, we briefly report on the configurational properties of polymers with more than one active block. We choose to report these data as a function of the minimum contour distance between the beginning of the closest active section and the head of the polymer, for the sake of coherence with the rest of the manuscript. 

\begin{figure*}[ht]
	\centering
		\includegraphics[width=5.8cm]{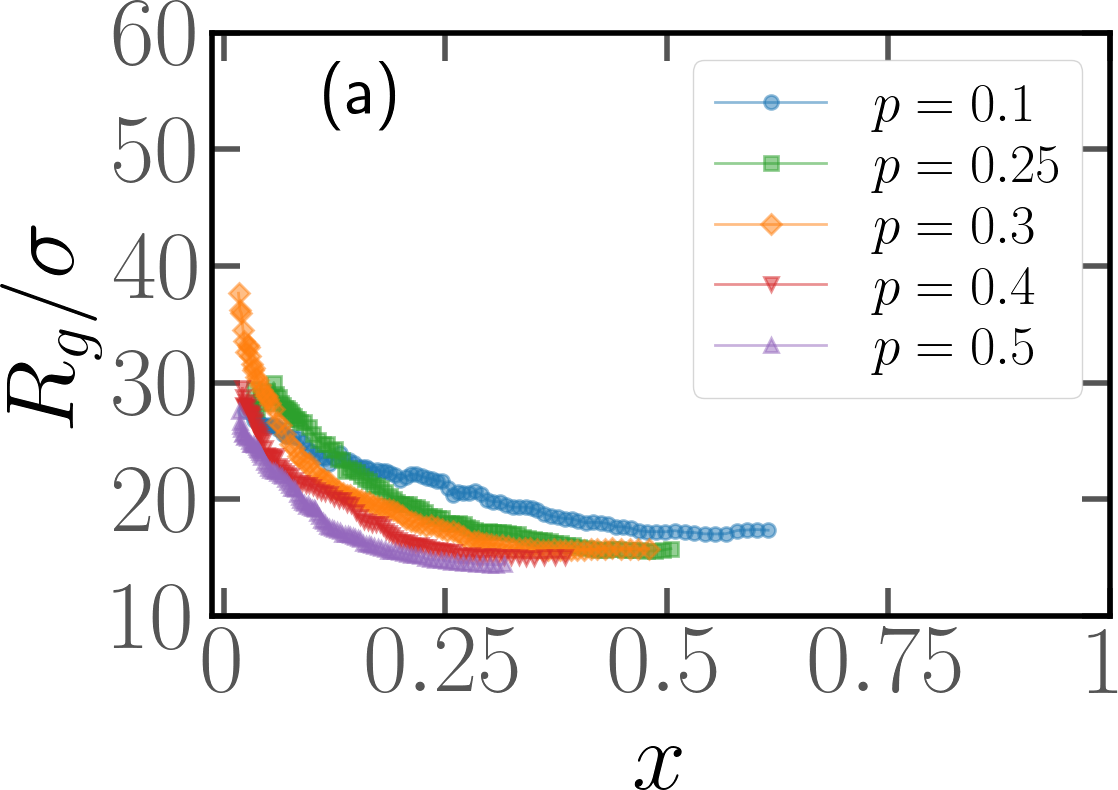}
		\includegraphics[width=5.8cm]{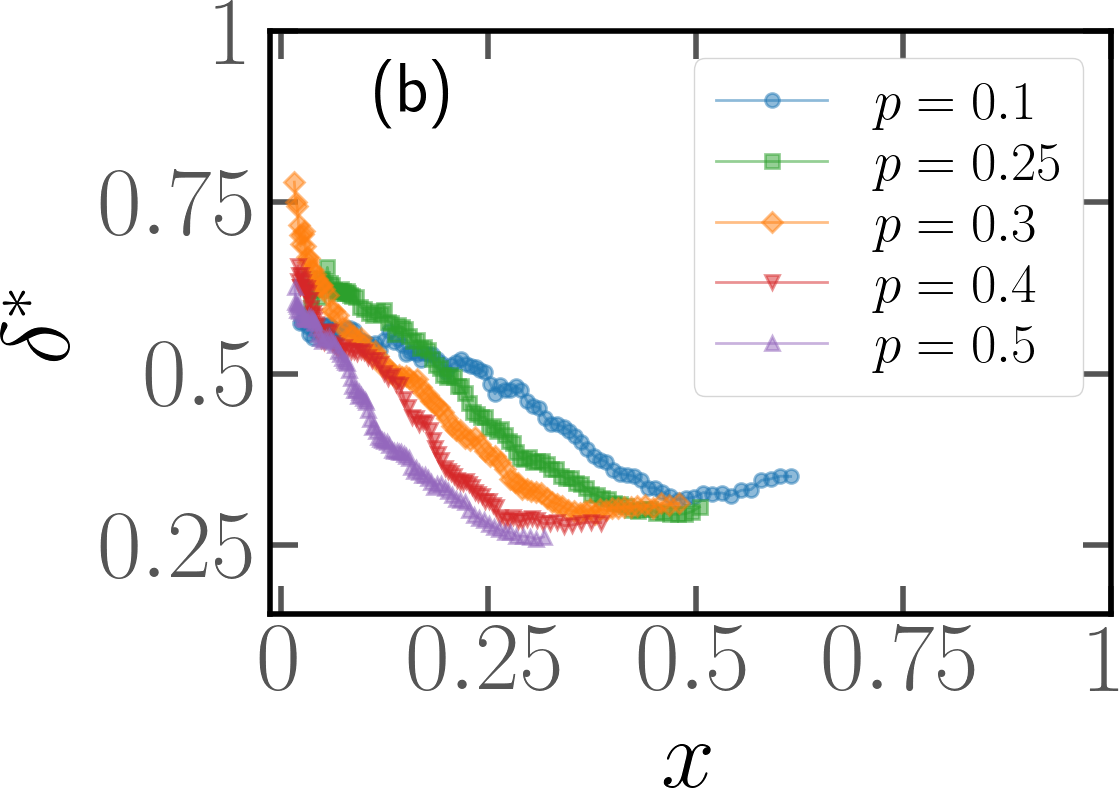}
		\includegraphics[width=5.8cm]{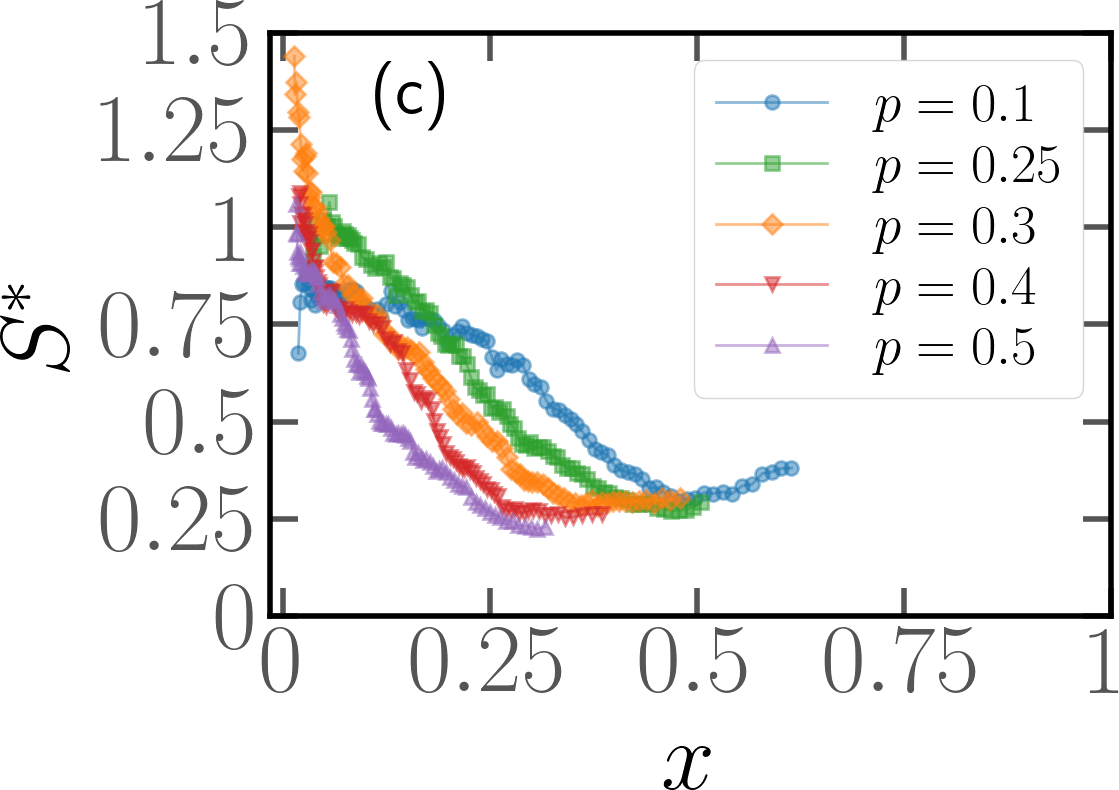}
		\includegraphics[width=5.8cm]{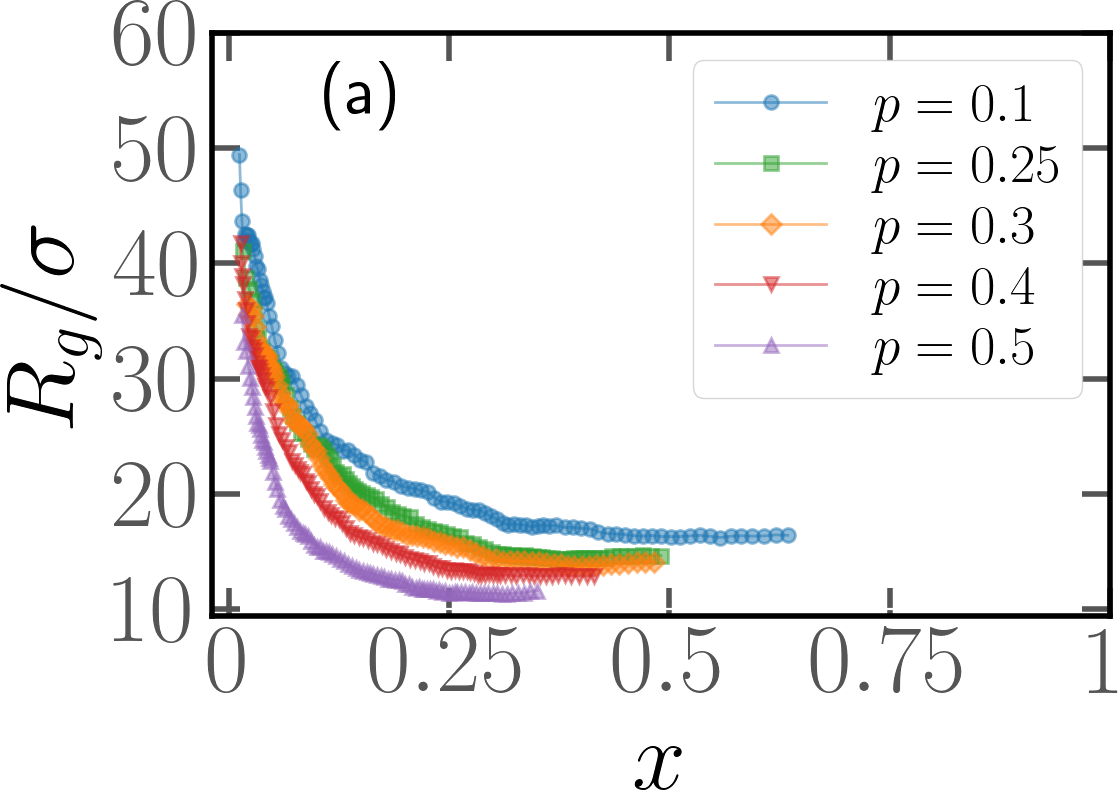}
		\includegraphics[width=5.8cm]{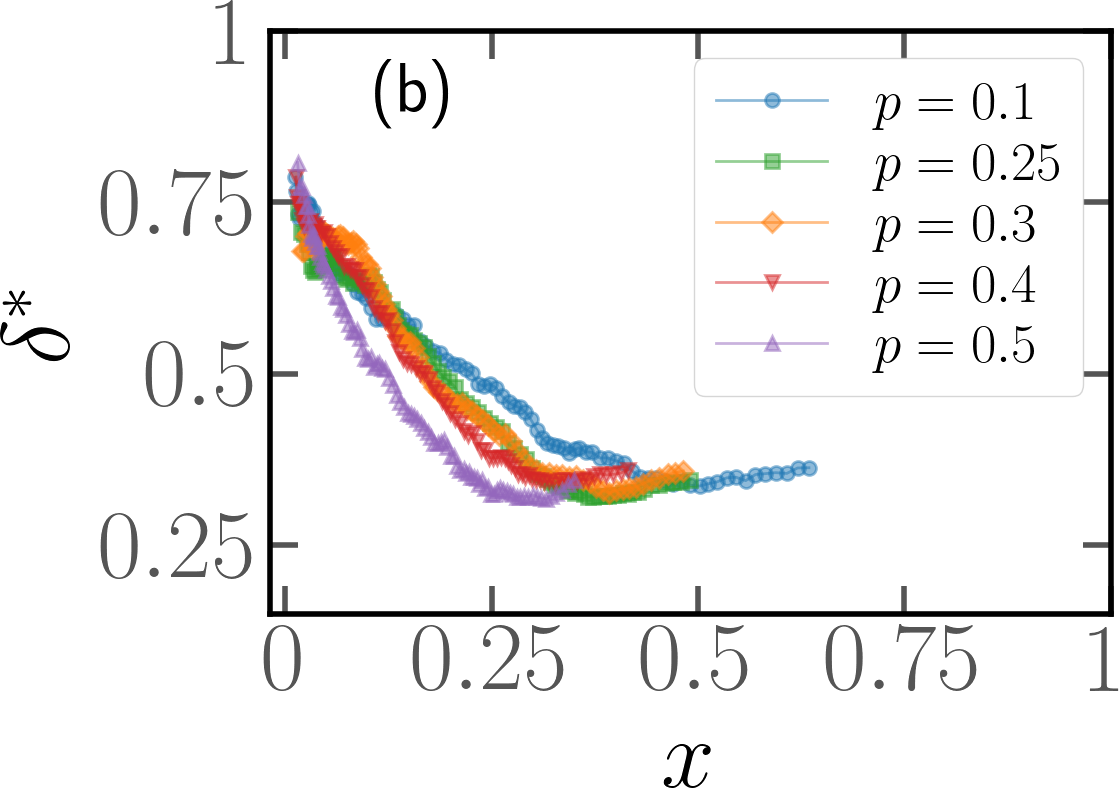}
		\includegraphics[width=5.8cm]{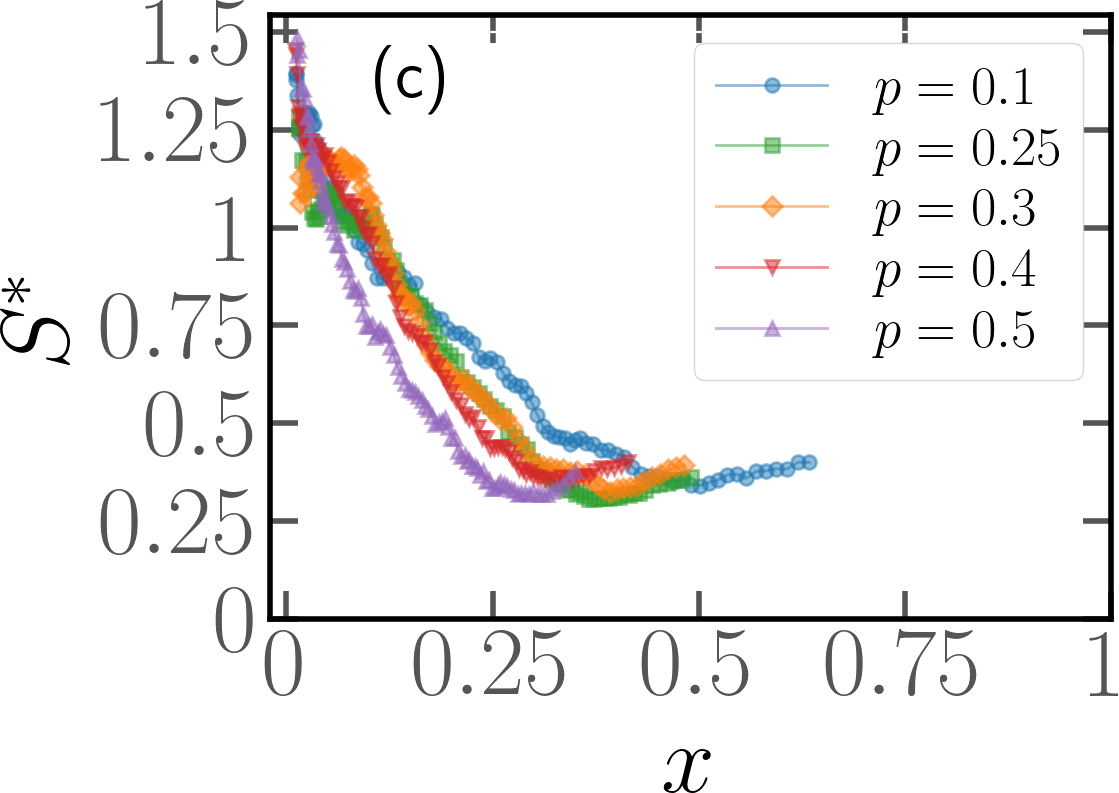}
	\caption{Mean Gyration Radius (left), asphericity (middle), and prolatness (right) as a function of the parameter $x$, for polymer chains of $N=$500 monomers with two non-overlapping active blocks, $Pe=0.1$ (top row) and $Pe=10$ (bottom row) and different values of $p$.}
	\label{fig:geo_q_2}
\end{figure*}

In Fig.~\ref{fig:geo_q_2} we show the mean gyration radius, asphericity and prolateness as a function of the parameter $x$ for polymers of length $N=$500 with two non-overlapping active blocks  at $\mathrm{Pe}=$0.1 (top row), $\mathrm{Pe}=$10. (bottom row) and different values of $p$. We observe that the shape parameters retain the non-monotonic character also in presence of a second active block; however asphericity and prolateness appear to be more similar, upon varying the fraction of the active monomers $p$, with respect to the single block case. On the contrary, the gyration radius as a function of $x$ changes qualitatively in presence of a second active block, losing its non-monotonic character. However, the special nature of arrangements with small $x$ is maintained.

\begin{figure*}[ht]
	\centering
		\includegraphics[width=5.8cm]{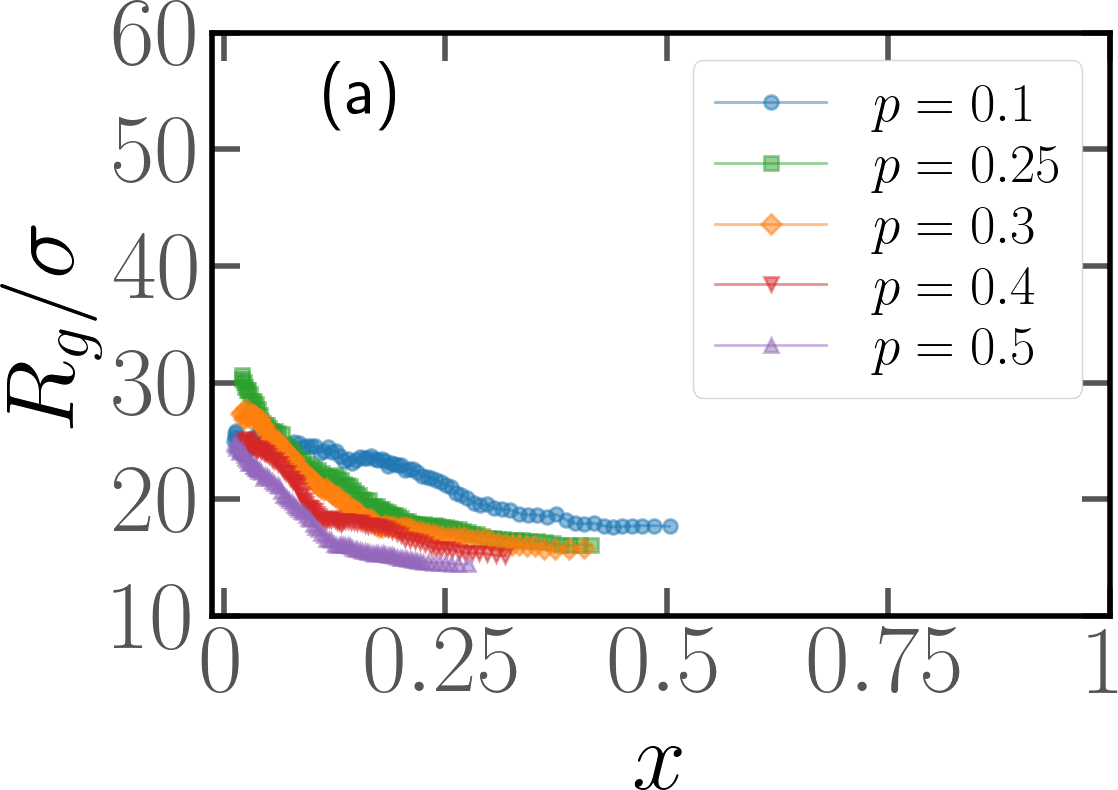}
		\includegraphics[width=5.8cm]{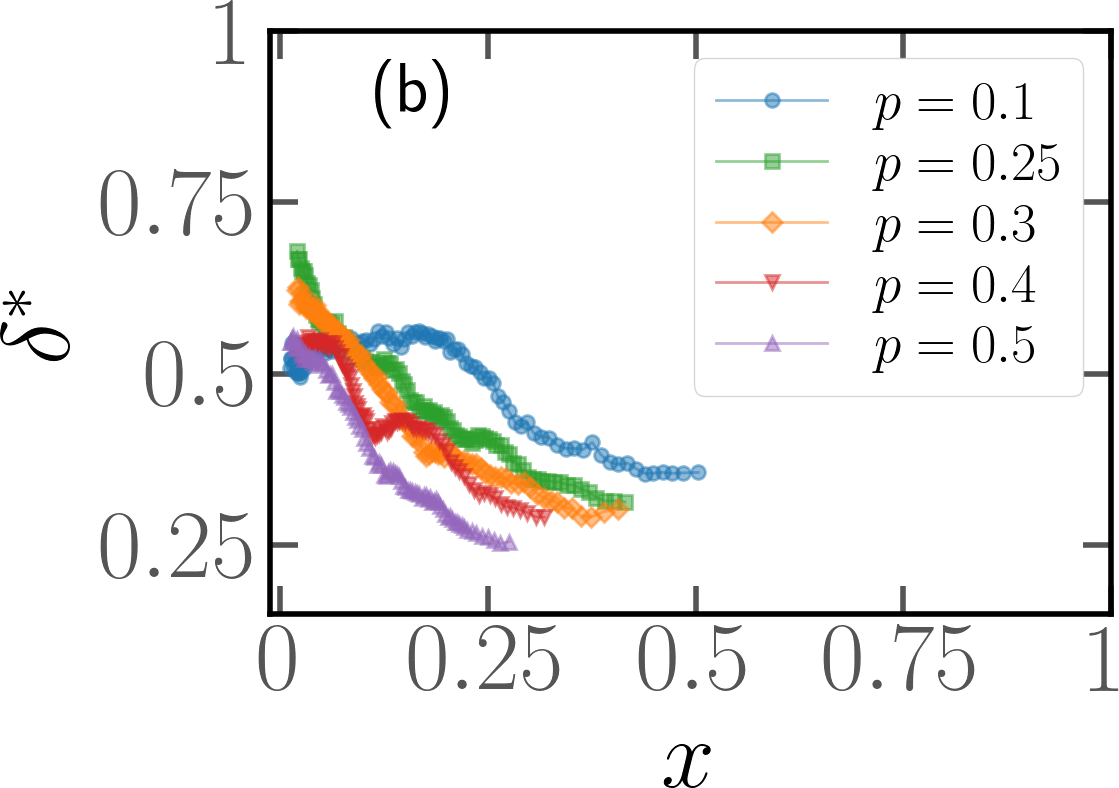}
		\includegraphics[width=5.8cm]{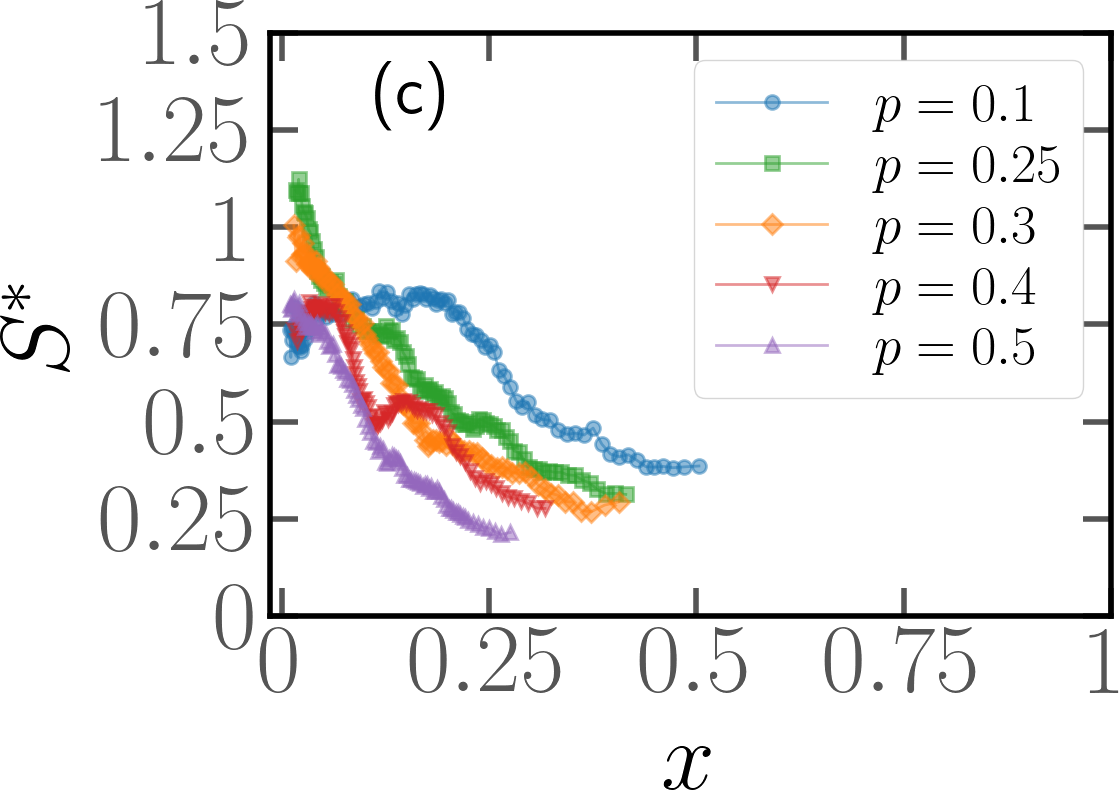}
		\includegraphics[width=5.8cm]{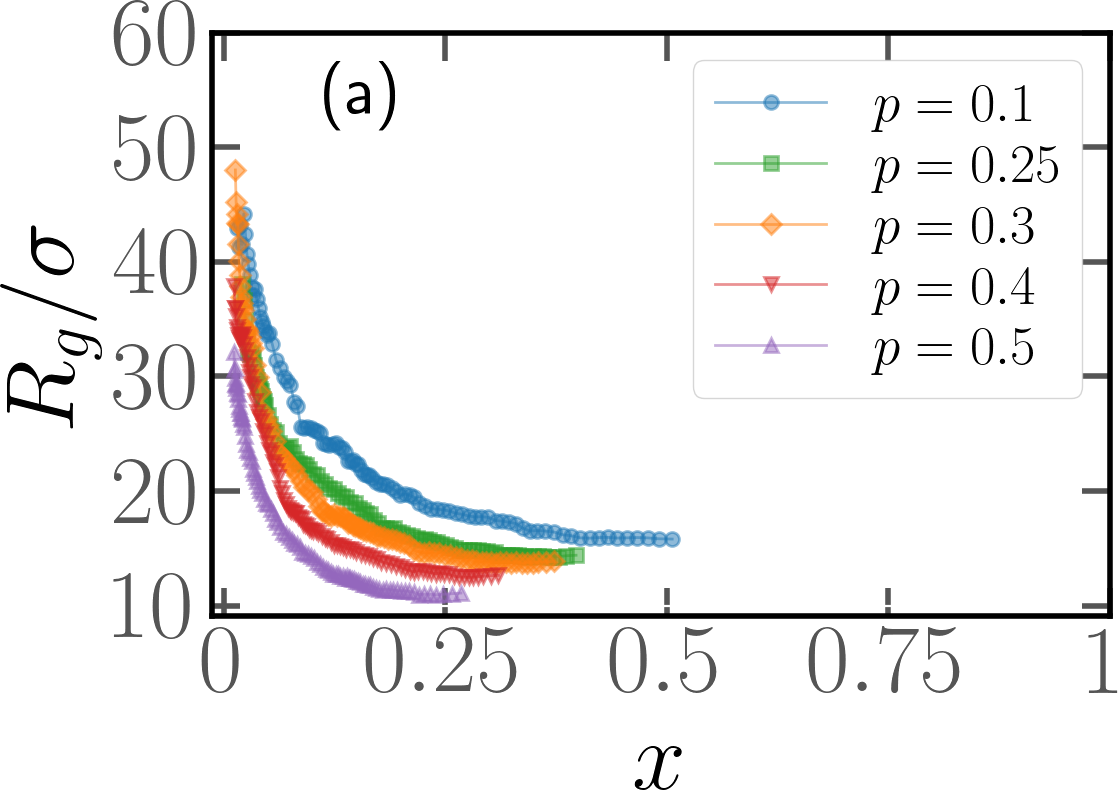}
		\includegraphics[width=5.8cm]{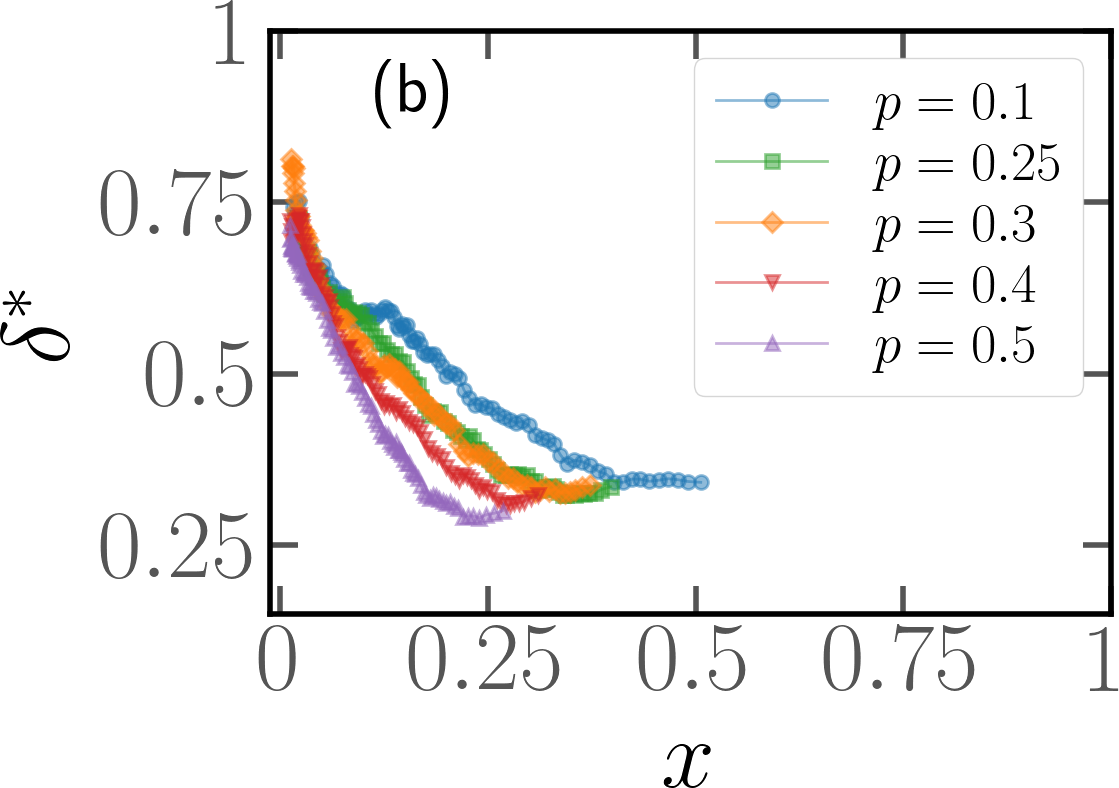}
		\includegraphics[width=5.8cm]{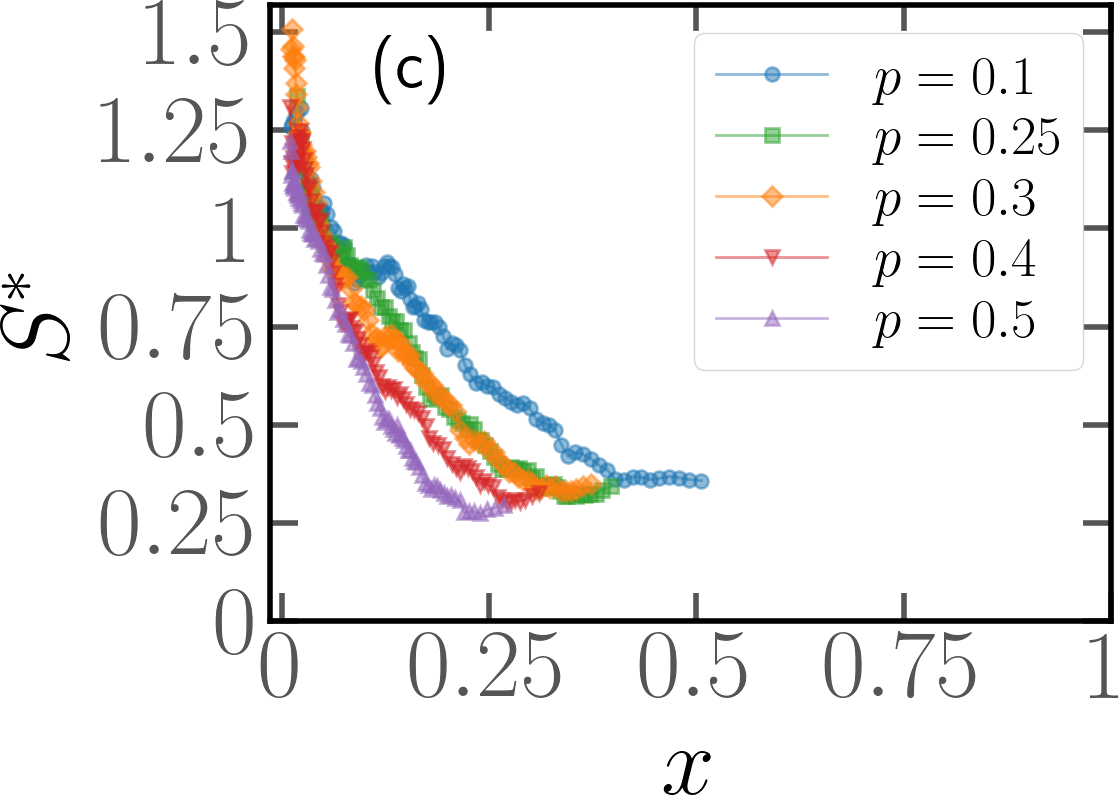}
	\caption{Mean Gyration Radius (left), asphericity (middle), and prolatness (right) as a function of the parameter $x$, for polymer chains of $N=$500 monomers with three non-overlapping active blocks, $Pe=0.1$ (top row) and $Pe=10$ (bottom row) and different values of $p$.}
	\label{fig:geo_q_3}
\end{figure*}

Finally, in Fig.~\ref{fig:geo_q_3} we show the mean gyration radius, asphericity and prolateness as a function of the parameter $x$ for polymers of length $N=$500 with three non-overlapping active blocks at $\mathrm{Pe}=$0.1 (top row), $\mathrm{Pe}=$10. (bottom row) and different values of $p$. We can make essentially the same observation as in the two blocks case. All in all, upon increasing the number of blocks, the parameter $x$ appears to become less effective, as highlighted by the more pronounced overlap between the curves, visible in both Figs.~\ref{fig:geo_q_2},~\ref{fig:geo_q_3}. Moreover, there is a further symmetry breaking in this representation as, for small values of $x$, multiple different arrangements share the same value of $x$, while the same happens at $x \to 1-p$ only to very few, very similar arrangements.  


\providecommand*{\mcitethebibliography}{\thebibliography}
\csname @ifundefined\endcsname{endmcitethebibliography}
{\let\endmcitethebibliography\endthebibliography}{}

\end{document}